\documentclass[letterpaper, oneside, 12pt]{article}
\usepackage{amsmath}
\usepackage{graphicx}%
\usepackage{amsfonts}%
\usepackage{amssymb}
\usepackage{overpic}
\usepackage[ruled]{algorithm}
\usepackage[noend]{algpseudocode}
\usepackage{subfig}
\usepackage{rotating}
\usepackage{epstopdf}
\usepackage{url}
\usepackage[table]{xcolor}
\usepackage{color}
\usepackage{comment}
\usepackage{enumerate}
\usepackage{mathtools}
\usepackage{multirow}
\usepackage{natbib}
\usepackage{pst-solides3d}
\usepackage{soul}
\usepackage{booktabs}

\usepackage{graphicx}
\usepackage{subfig}
\usepackage{algorithm}
\usepackage{multirow}
\usepackage{booktabs}
\usepackage{tabularx}
\usepackage{setspace}

\usepackage{colortbl}
\usepackage{makecell}
\usepackage{array}
\usepackage{comment}

\usepackage{setspace}
\usepackage[hmargin=1.0in,vmargin=1.0in]{geometry}

\usepackage[compact]{titlesec}
\titlespacing{\section}{0pt}{*0.8}{*0.8}
\titlespacing{\subsection}{0pt}{*0.8}{*0.8}
\titlespacing{\subsubsection}{0pt}{*0.8}{*0.8}

\newcommand{\bB}{ {\boldsymbol B} }

\newcommand{\bI}{ {\boldsymbol I} }

\newcommand{\bL}{ {\boldsymbol L} }

\newcommand{\bO}{ {\boldsymbol O} }

\newcommand{\bs}{ {\boldsymbol s} }

\newcommand{\bw}{ {\boldsymbol w} }

\newcommand{\bx}{ {\boldsymbol x} }

\newcommand{\bY}{ {\boldsymbol Y} }

\newcommand{\balpha}{ {\boldsymbol \alpha} }

\newcommand{\bbeta}{ {\boldsymbol \beta} }

\newcommand{\bgamma}{ {\boldsymbol \gamma} }

\newcommand{\bdelta}{ {\boldsymbol \delta} }

\newcommand{\bLambda}{ {\boldsymbol \Lambda} }

\newcommand{\bSigma}{ {\boldsymbol \Sigma} }
\newcommand{\btheta}{ {\boldsymbol \theta} }

\newcommand{\bzero}{ {\boldsymbol 0} }

\title{Integrative Predictor-Dependent Learning of Network Data and Spatially Correlated Nodal Attributes for Multimodal Brain Imaging in Aging}

\author{Jose Rodriguez-Acosta\footnotemark[1]
    \and 
    Sharmistha Guha\footnotemark[1]
    \and
    Jessica Bernard\footnotemark[2]
    \and
    Thamires Magalhaes\footnotemark[2]
    \and
    Kaitlin McOwen\footnotemark[2]
}

\date{}

\begin{document}

\maketitle

\footnotetext[1]{Department of Statistics, Texas A\&M University}

\footnotetext[2]{Department of Psychological and Brain Sciences, Texas A\&M University}

\begin{abstract}
This article introduces a predictor-dependent joint modeling framework for network data obtained from multiple subjects over a \emph{shared set of nodes with spatial co-ordinates} and spatially correlated attributes defined at these nodes. The proposed framework is highly flexible, allowing for concurrent inference on nodes significantly associated with a predictor, spatial associations of nodal attributes across nodes, and the regression relationship between a predictor and an edge connecting a pair of nodes or a specific nodal attribute. Empirical results, as demonstrated in simulation studies, indicate a superior performance of the proposed approach due to accounting for network structure and spatial correlation in the data simultaneously. The proposed methodology analyzes multimodal brain imaging data collected first-hand in the coauthor's Lifespan Cognitive and Motor Neuroimaging Laboratory, with a focus on integrating structural and functional information. It examines brain connectivity, represented as a connectome network across regions of interest (ROIs) derived from functional magnetic resonance imaging (fMRI), while simultaneously incorporating ROI-specific attributes obtained from structural MRI data, for each subject. Subject-specific aging-related features and the spatial locations of each ROI are also incorporated into the analysis. This comprehensive framework facilitates robust inference on the associations between predictors and brain connectivity patterns, the spatial relationships among ROI-specific attributes, and the regression relationships involving edges or ROI-specific attributes with aging-related predictors. By integrating these diverse data sources, the approach provides a deeper understanding of the complex interplay between brain structure, function, aging-related changes, and external predictors.
As a model-based Bayesian approach, it provides uncertainty quantification for all inferences, offering robust and reliable results, particularly in scenarios with limited sample size.
\end{abstract}


\newpage

\section{Introduction}

This article develops a unified predictor-dependent learning framework designed to jointly analyze networks and spatially correlated nodal attributes across multiple subjects. The motivation for this methodological advancement arises from a clinical study focused on understanding brain function and structure associated with aging in subjects. The study collects multi-modal neuroimaging data for each participant, including (a) \emph{functional brain network information} and (b) \emph{brain structural information}. The functional brain network information considers each region-of-interest (ROI) as a network node and quantifies connectivity (operationalized as the Pearson's correlation across timeseries) between pairs of network nodes using the blood oxygen-level dependent (BOLD) signal collected at rest (absence of task). This is measured using functional magnetic resonance imaging (fMRI). Brain structural information is derived from structural MRI (sMRI) and is represented as a ROI or node-specific attribute, termed the \emph{structural attribute}, which displays spatial correlations across ROIs. Spatial co-ordinates of ROIs are also available for the study.
Thereafter, using a common brain atlas for all subjects, we segment the human brain into different ROIs. The study integrates information from both brain network connectivity and brain structural attributes to achieve two primary objectives: \textbf{(o1)} draw inference on brain ROIs significantly related to aging; and \textbf{(o2)} estimate the regression relationship between the network and structural attributes with each predictor. 

Broadly, the literature addresses the joint modeling of networks and nodal attributes through three distinct approaches. One approach focuses on modeling the network structure conditional on nodal attributes, capturing how relationships among nodes are formed based on these attributes, a process referred to as ``selection.'' Models of selection are typically constructed by regressing a network edge on node-specific attributes, employing tools such as Exponential Random Graph Models (ERGMs) \citep{Holland_1981, robins2007recent} or mixed-effects generalized linear models \citep{Wasserman_1987, Holland_1983, hoff2002latent, hoff2005bilinear}. Another approach involves models of nodal attributes and their association, conditional on the network structure. These models seek to understand how relationships impact the attributes of nodes in a network, a phenomenon referred to as ``influence'' or ``contagion.'' 
Models of contagion typically involve regressing nodal attributes on those of other nodes in the network, with methods including simultaneous autoregressive (SAR) models \citep{lin2010identifying} and threshold models \citep{watts2009threshold}, among others  \citep{christakis2007spread, fowler2008dynamic, shoham2015modeling}. Since distinguishing between selection and contagion is a challenging problem, a third set of approaches simultaneously models both networks and nodal attributes.
These joint models often assume a bilinear representation of network edges with node-specific latent effects, either modeling these latent effects alongside nodal attributes \citep{fosdick2015testing}, or utilizing a shared set of latent effects \citep{guhaniyogi2020joint, zhang2022joint} to jointly model the network and nodal attributes.

Our approach diverges from existing literature on joint modeling of network and nodal attributes in several critical aspects. First, the existing literature is predominantly unsupervised, meaning that the joint modeling of networks and nodal attributes often involves predictors at the \emph{node} or \emph{edge} level, but typically omits \emph{subject-level} predictors.
Additionally, the primary objective of this literature revolves around drawing inferences on whether the network and nodal attributes are independent across all nodes. For instance, \cite{fosdick2015testing} develop a likelihood ratio testing framework to determine independence between the network and nodal attributes at all nodes. \cite{guhaniyogi2020joint} develop a Bayesian testing framework to examine independence between the network and nodal attributes when they jointly co-evolve over time.
In contrast, our inferential focus is distinct, centered on making inferences regarding network nodes significantly associated with a subject-level predictor of interest, jointly learned from both the network and nodal attributes. Thus, while prior studies have often focused on statistical testing or inference regarding network and nodal attribute relationships, this article addresses a different inferential goal.
Second, since much of the existing literature in this domain primarily draws motivation from social networks, the spatial coordinates of the nodes are generally not available. Consequently, the existing literature does not take into account spatial associations between nodal attributes in the joint modeling of network and nodal attributes.

This article addresses these identified gaps in the current literature by introducing a novel Bayesian framework that jointly models network structures and nodal attributes as responses to subject-level predictors. The corresponding regression coefficients for subject-level predictors are termed \emph{network coefficients} and \emph{structural coefficients}, respectively. 
To address inferential goals \textbf{(o1)} and \textbf{(o2)}, we develop a Bayesian hierarchical model with a joint prior linking these coefficients. A low-rank structure is used for network coefficients, representing each edge as a bilinear function of  node-specific latent variables connecting the edge. The goal \textbf{(o1)} of identifying influential network nodes is embedded within a nonlinear variable selection framework wherein a node-specific latent variable and the structural coefficient for the specific node are jointly assigned a spike-and-slab prior. The spike-and-slab prior involves node-specific latent activation indicators, which are shared among both sets of coefficients, and are instrumental in identifying influential nodes. Specifically, when a node's activation indicator is zero, both its attribute and all connected edges are unrelated to the predictor, marking it \emph{uninfluential}. This shared prior also allows information borrowing between the two response types. To capture spatial correlation among nodal attributes (relevant for \textbf{(o2)}), we introduce Gaussian process priors for spatial random effects in the regression model for the nodal attributes. The proposed framework facilitates efficient Bayesian computation, enables precise uncertainty quantification in identifying influential nodes crucial for the study of aging with multi-modal neuroimaging data, accounts for spatial correlation between nodal attributes, and produces well-calibrated interval estimates for regression coefficients.

While our method is designed to achieve inferential goals \textbf{(o1)} and \textbf{(o2)} through predictor-dependent joint modeling of networks and nodal attributes, it also contributes substantially to the literature of Bayesian regression with multi-modal image responses having diverse structures. Existing approaches typically view images as objects, such as tensors, functions, or networks, and focus on single-object response regression \citep{zeng2024bayesian, guha2021bayesian}. In the presence of multiple images, the existing literature often ignores relationships between multiple objects and their shared structure across modalities, which can lead to inaccurate and noise-sensitive inference \citep{CALHOUN2016230}. Recent literature has emphasized joint modeling of multi-modal imaging data, which has led to the literature on joint regression of multiple images. For example, recent work in joint structural-functional neuroimaging adopts this framework to regress multiple brain imaging outcomes on cognitive or clinical measures, thereby identifying common and modality-specific patterns associated with disease, aging, or behavior in a statistically principled manner \citep{gutierrez2025multiobject}. Parallel to these statistical approaches, generative deep learning methods have attracted increasing attention for multimodal neuroimaging. Such models seek to synthesize structural or functional images conditional on covariates (e.g., diagnosis, age) or other imaging modalities, using architectures such as conditional GANs, and variational autoencoders \citep{suzuki2017overview, bowles2018gan, wolterink2017deep}. These tools offer powerful predictive and simulation capabilities, but often provide limited interpretability and formal uncertainty quantification. In fact, to the best of our knowledge, there is still a lack of a Bayesian framework that jointly regresses network and nodal attributes on predictors, with a goal of addressing the key objective of node identification,  while incorporating spatial correlation among nodal features.

The subsequent sections of the article are organized as follows. Section~\ref{sec:data_description} describes the motivating multi-modal neuroimaging dataset on aging. Section~\ref{sec:model_and_prior} presents the model development and prior structures to learn the predictor-dependent association between the network and nodal attributes. The computation of the posterior distribution is discussed in Section~\ref{sec:posterior_comp}. Empirical investigations of the model under various simulation scenarios are provided in Section~\ref{sec:simulations}. Section~\ref{sec:realdata} describes the joint analysis of fMRI and sMRI data, demonstrating the effective performance of our approach compared to its competitors. Lastly, Section~\ref{sec:conclusion} concludes the article. 

\section{Multi-modal Neuroimaging Data on Aging}\label{sec:data_description}

\subsection{Data Description}
We use the proposed methodology to analyze multi-modal human neuroimaging data collected first-hand by co-author Bernard and colleagues within the
\textbf{Lifespan Cognitive and Motor Neuroimaging Laboratory} at Texas A\&M University, College Station.  These data are from a sample of 138 healthy, independently living adults between the ages of 35 and 86 years (mean age: 57 years; 54\% female).
All participants were recruited as part of a longitudinal study of brain and behavior across adulthood and aging.  After exclusions for incomplete data or noise, we were left with a sample of 126 individuals for the analyses conducted here. As part of this  longitudinal study, all participants completed behavioral and neuroimaging data collection sessions. 
Exclusion criteria were the following: a history of neurological disease, stroke, or formal diagnosis of psychiatric illness (e.g., depression or anxiety), contraindications for the brain imaging environment (metal implants in the  body, claustrophobia, inability to lay flat for an extended period of time), and the use of hormone therapy (HTh) or hormonal contraceptives including intrauterine devices (IUDs), possible use of continuous birth control (oral), and no history of hysterectomy in the past 10 years.  

Participants underwent structural and resting-state MRI using a Siemens Magnetom Verio 3.0 Tesla scanner with a 32-channel head coil. 
Structural MRI included a high-resolution T1-weighted 3D magnetization prepared rapid gradient multi-echo (MPRAGE) scan (TR = 2400 ms; acquisition time = 7 minutes; voxel size = 0.8 mm\textsuperscript{3}) and a high-resolution T2-weighted scan (TR = 3200 ms; acquisition time = 5.5 minutes; voxel size = 0.8 mm\textsuperscript{3}), both with a multiband acceleration factor of 2. 
Resting-state imaging comprised four blood-oxygen-level-dependent (BOLD) functional connectivity (fMRI) scans with a multiband factor of 8, 488 volumes, TR of 720 ms, and 2.5 mm\textsuperscript{3} voxels.  
Scanning protocols were adapted from the Human Connectome Project \citep{Harms_2018} and the Center for Magnetic Resonance Research at the University of Minnesota to ensure future data sharing and reproducibility. 

All preprocessing steps follow the recent work on this dataset as described in \cite{Hicks_2023} and \cite{Ballard_2022}. 
The images were converted from DICOM to NIFTI format and organized according to the Brain Imaging Data Structure (BIDS, version 1.6.0) using the 
bidskit docker container (\url{https://github.com/jmtyszka/bidskit}). A single volume was extracted from two oppositely coded BOLD images to estimate $B0$ field maps using the split tool from the FMRIB Software Library (FSL) package \citep{Jenkinson_2012}. Subsequently, anatomical and functional images were preprocessed using fMRIPrep (
 \url{https://fmriprep.org/}), which involves a series of automated procedures to prepare fMRI data for further analysis. This includes aligning the functional volume with the anatomical image for accurate spatial registration, correcting motion to account for participant movement during image acquisition, correcting fieldmap distortions caused by the non-uniform magnetic field, segmenting the anatomical image into distinct tissues (e.g., gray matter, white matter, cerebrospinal fluid (CSF)), stripping the skull from the anatomical image to improve segmentation quality and reduce artifacts, normalizing the data to a common space for cross-participant or cross-study comparisons, aligning motion-corrected functional volumes with the normalized anatomical image for functional-anatomical integration, and applying spatial smoothing to enhance the signal-to-noise ratio and reduce artifacts. We continued the rest of the analyses using the CONN toolbox (version 21a) \citep{WhitfieldGabrieli2012ConnAF}. 

\begin{figure}
\centering
\includegraphics[width=\linewidth]{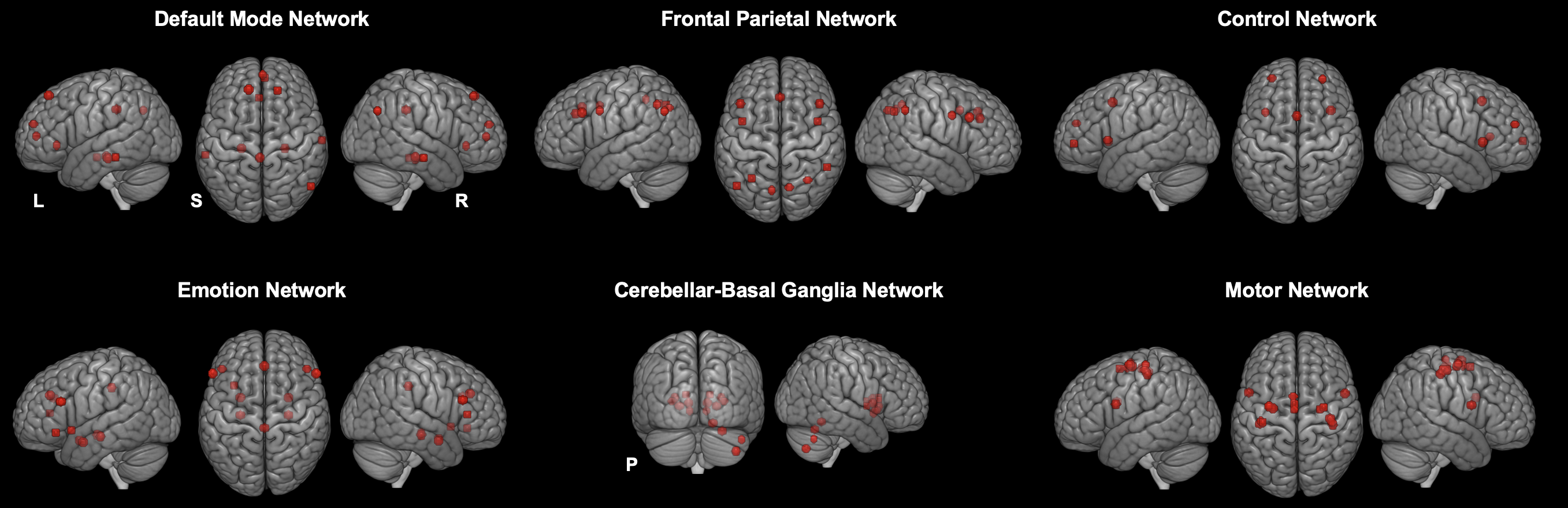}
\caption{Figure illustrates the whole-brain network, highlighting the subnetworks through red nodes.}
\label{sim-circos-fig-4}
\end{figure}

\noindent\underline{\textbf{Brain network data.}} 
Regions of interest (ROIs) were defined based on well-known cortical networks, following the networks used in \cite{Jackson_2023}, which included the \emph{default mode network, frontal-parietal network, control network, emotion network, motor network}, and a subcortical \emph{cerebellar-basal ganglia network}.
This resulted in a total of 75 ROIs. However, to ensure alignment with structural segmentations (described below), 6 ROIs were excluded, resulting in a final set of 69 ROIs. 
These 69 ROIs within the cortical and subcortical networks are marked in red in Figure~\ref{sim-circos-fig-4}.  
For each participant we extracted a time series for each of the 69 ROIs, and they were cross-correlated with one another using bivariate analyses. The result was a correlation matrix for each participant. As with the preprocessing above, this was completed using the CONN toolbox (version 21a) \citep{WhitfieldGabrieli2012ConnAF}.

\noindent\underline{\textbf{Structural data.}} 
Alongside the resting-state correlation matrices, we also quantified brain structural information associated with each ROI.  
Functional MNI coordinates from \cite{Jackson_2023} used in the analyses described above were matched to structural regions through the use of MRIcroGL and the AAL atlas  \citep{TZOURIOMAZOYER2002273}. 

\noindent\underline{\textbf{Subject-level attributes.}} 
All participants completed the Purdue Pegboard task \citep{lawson2019purdue}, which allows quantification of fine motor function and dexterity. This task uses a pegboard with two columns of 25 holes and a tray with four cups (pegs, cylinders, washers, pegs). Participants complete four conditions: \textbf{Unimanual (Right \& Left):} place pegs into one column using one hand; each hand tested separately for 30 seconds; \textbf{Bimanual:} place pegs into both columns simultaneously using both hands for 30 seconds; \textbf{Assembly:} Assemble peg-washer-cylinder-washer using alternating hands for 60 seconds.
Each condition is repeated four times. Performance is measured by counting pegs placed (unimanual), pairs completed (bimanual), and total pieces assembled (assembly), then averaged across repetitions.
The behavioral attribute is the total score across all conditions of the Purdue Pegboard test for each participant, referred to as the \emph{Aggregate Pegboard Score (APS).}

\noindent \underline{\textbf{Scientific goals of the data analysis.}}
The primary scientific objective of this study is to identify brain regions of interest (ROIs) that are significantly linked to cognitive and motor aging. To achieve this, we represent Z-transformed correlation matrices across all ROIs as a brain network and treat the structural volumes of all ROIs as spatially-varying, node-specific outcomes. We then develop a novel statistical methodology (described in the following sections) to capture their predictor-specific (Aggregate Pegboard Score) joint dependence, accounting for race, sex, and age. The proposed method enables identification of regions that most strongly predict motor performance, while also quantifying the statistical uncertainty of these findings.

\section{Model Formulation and Prior Structure}\label{sec:model_and_prior}

Assume that the data is obtained over $n$ subjects, and let $\mathcal{G}_i$ denote the network data for the $i$th subject, $i=1,...,n$. We assume that the set of networks $\{\mathcal{G}_i:i=1,..,n\}$ is defined over a common set of nodes denoted by $\mathcal{N}=\{\mathcal{N}(\bs_1),...,\mathcal{N}(\bs_V)\}$, where $|\mathcal{N}|=V$ represents the number of nodes and $\bs_1,...,\bs_V$ are the spatial locations of the nodes. The network $\mathcal{G}_i$ is represented by a $V\times V$ adjacency matrix $\bY_i\in\mathbb{R}^{V\times V}$, where the entry $y_{i}(u,v)$ at position $(u,v)$ signifies the strength of association between nodes $\mathcal{N}(\bs_u)$ and $\mathcal{N}(\bs_v)$ for the $i$th subject, $u,v=1,...,V$. This article concentrates specifically on undirected networks without self-relationships, a characteristic considered scientifically meaningful for the networks derived from fMRI data in Section \ref{sec:realdata}. This implies that the adjacency matrix $\bY_i$ is symmetric, and its diagonal entries are zero. Let $z_i(\bs_v)\in\mathbb{R}$ represent the attribute observed at node $\mathcal{N}(\bs_v)$ for subject $i$, $\bx_i\in\mathbb{R}^p$ denote the predictors of interest, and $\bw_i\in\mathbb{R}^q$ be the auxiliary predictors. In the context of the motivating scientific problem in Section \ref{sec:data_description}, race, age and sex are the auxiliary predictors, while the Aggregate Pegboard Score (APS) 
is the predictor of interest.

\subsection{A Joint Model for Network and Nodal Attributes Incorporating Spatial Information}\label{sec:model_formulation}
The following model is developed to effectively and flexibly characterize the relationship between the network, the predictors of interest $\bx_i$, and auxiliary predictors $\bw_i$:
\begin{align}
E[y_{i}(u,v)|\bx_i,\bw_i] = G(\mu_y+\sum_{k=1}^p\beta_k(u,v)x_{i,k} + \sum_{j=1}^q\gamma_{j,y}w_{i,j}),
\end{align}
where $\gamma_{1,y},...,\gamma_{q,y}$ are the coefficients for the auxiliary predictors, $\mu_y$ is the intercept and $\beta_k(u,v)$ is the coefficient determining the relationship between the $(u,v)$th edge and the $k$th predictor of interest. Here, $G(\cdot)$ is an appropriate link function corresponding to the network edge. For example, when network edges are continuous, we can set $G(\cdot)$ to be the identity link function, leading to
\begin{align}\label{network_equation}
y_{i}(u,v) = \mu_y+\sum_{k=1}^p\beta_k(u,v)x_{i,k} + \sum_{j=1}^q\gamma_{j,y}w_{i,j} + \epsilon_{i}(u,v), \:\:\epsilon_{i}(u,v)\sim N(0,\tau_y^2),
\end{align}
where $\tau_y^2$ is the error variance. The nodal attribute is assumed to be continuous and is spatially correlated over network nodes.
To account for its relationship with the covariates after accounting for spatial correlation, a series of conditionally independent models are proposed over $i=1,...,n$,
\begin{align}\label{structural_equation}
 z_{i}(\bs_v) = \mu_{z}+\sum_{k=1}^p\alpha_{k}(v) x_{i,k} + \sum_{j=1}^q\gamma_{j,z} w_{i,j} + \delta_{i}(\bs_v),
\end{align}
where $\mu_z$ is the intercept, $\alpha_k(v)$ is the coefficient corresponding to the predictors of interest, $\gamma_{1,z},...,\gamma_{q,z}$ are the coefficients for the auxiliary predictors, and $\delta_i(\bs_v)$ is the spatially dependent random effect accounting for the correlation structure of the nodal attribute over the nodes.

\subsection{Prior Structure}\label{sec:prior_structure}
In this section, we describe the construction of the prior structure on model coefficients, with particular emphasis on coefficients $\bbeta=\{\beta_k(u,v):k=1,..,p; 1\leq u\neq v\leq V\}$, $\balpha=\{\alpha_k(v):k=1,...,p; v=1,...,V\}$, and functions $\{\delta_i(\cdot):i=1,..,n\}$. To this end, we formulate a joint prior for the coefficients $\bbeta=\{\beta_k(u,v):k=1,..,p; 1\leq u\neq v\leq V\}$ and $\balpha=\{\alpha_k(v):k=1,...,p; v=1,...,V\}$, establishing a predictor-dependent association between the network and the nodal attribute, while taking into account the network's topology. The construction of the joint prior on $\bbeta$ and $\balpha$ addresses: (a) identification of nodes associated with a predictor, incorporating uncertainty; and (b) efficient computation of the posterior based on the proposed prior. Furthermore, a prior structure is imposed on functions $\{\delta_i(\cdot):i=1,..,n\}$ which is crucial for introducing spatially dependent correlation among the attribute over nodes. A detailed construction of prior distributions for $\bbeta$, $\balpha$ and $\{\delta_i(\cdot):i=1,..,n\}$ is provided below.

\noindent\underline{\textbf{Joint prior construction on coefficients $\bbeta$ and $\balpha$.}} In the construction of a joint prior on $\bbeta$ and $\balpha$, we first introduce a low-rank structure of the coefficient $\beta_k(u,v)$ as
\begin{equation}\label{param}
\beta_k(u,v)=\sum_{r=1}^R\lambda_{k,r}\theta_{k,r}(u)\theta_{k,r}(v),\:\beta_k(u,u)=0,\:\:1\leq u\neq v\leq V.
\end{equation}
Here $\btheta_k(v)=(\theta_{k,1}(v),...,\theta_{k,R}(v))^T\in\mathbb{R}^{R}$, for $k=1,...,p$, is a collection of $R$-dimensional latent variables, one for each node and each predictor of interest, and $\lambda_{k,r}\in\{-1,0,1\}$ determines if the $r$th summand in (\ref{param}) is relevant in model fitting. Since the choice of $R$ is arbitrary, allowing $\lambda_{k,r}$ to be $0$ protects the model from over-fitting. The $\lambda_{k,r}$'s are assigned a Multinomial-Dirichlet prior given by,
\begin{align}
&\lambda_{k,r}\sim \pi_{k,1,r}I(\lambda_{k,r}=0)+\pi_{k,2,r}I(\lambda_{k,r}=1)+\pi_{k,3,r}I(\lambda_{k,r}=-1),\nonumber\\
&(\pi_{k,1,r},\pi_{k,2,r},\pi_{k,3,r})\sim Dirichlet(r^{\xi},1,1),\:\:\:  \xi> 1.
\end{align}

The low-rank formulation in (\ref{param}) is motivated by multiple considerations, including the well-documented presence of transitivity effects in network data  \citep{RevModPhys.74.47}. Specifically, if the edges between node $\mathcal{N}(\bs_u)$ and both $\mathcal{N}(\bs_v)$ and $\mathcal{N}(\bs_{v'})$ are strongly associated with the $k$th predictor, then, due to transitivity, it is likely that the edge between $\mathcal{N}(\bs_v)$ and $\mathcal{N}(\bs_{v'})$ is also influential. Such patterns of interdependence among edges are commonly observed in real-world networks. A low-rank structure on the coefficient matrix $\bB_k = ((\beta_k(u,v)))_{u,v=1}^V$ naturally accommodates these effects. By modeling $\bB_k$ as a product of low-dimensional latent factors, the model implicitly assumes that edge influences are driven by node-specific latent traits. This induces a shared structure across rows and columns of $\bB_k$, leading to correlated patterns in edge importance and capturing the transitivity often exhibited in complex networks \citep{rodriguez2025supervised}.

Moreover, the low-rank formulation allows direct inference on network nodes through the node-specific latent vectors $\btheta_k(1),...,\btheta_k(V)$, which can be interpreted as the positions of the nodes in a latent space, with the strength of association or edge between nodes $\mathcal{N}(\bs_u)$ and $\mathcal{N}(\bs_v)$ being controlled by the inner product or the angular distance between the vectors $\btheta_k(u)$ and $\btheta_k(v)$. The assumed low-rank structure additionally offers parsimony by reducing the number of estimable parameters from $pV(V-1)/2$ to $pVR$, typically with $R\ll V$.

Depending on $\lambda_{k,r}$'s, the node specific latent variables $\btheta_k(v)$'s may become unidentifiable. For example, when $\lambda_{k,r}=1$ for all $r=1,..,R$, $\beta_k(u,v)=\btheta_k(u)^T\bLambda_k\btheta_k(v)=(\bO\btheta_k(u))^T\bLambda_k(\bO\btheta_k(v))$, for any orthogonal matrix $\bO$. While this implies that posterior inference on $\btheta_k(v)$'s may not always be meaningful, our focus is on the event $\{\btheta_k(v)=\bzero,\alpha_k(v)=0\}$ for each $k$, which is identifiable (since $\bzero$-valued latent vectors are invariant to orthogonal transformations) and is critical to drawing inference on the nodes related to the $k$-th predictor of interest. 
Indeed, $\btheta_k(v)$ and $\alpha_k(v)$ capture the strength of association between the $k$th predictor of interest and the network structure and nodal attributes, respectively.

In particular,
the $k$th predictor of interest $x_{i,k}$ is unrelated to the node $\mathcal{N}(\bs_v)$, if both $\btheta_k(v)=\bzero$ and $\alpha_k(v)=0$. We view the problem of identifying influential nodes as a high-dimensional variable selection problem and formulate variable selection prior distributions on $\btheta_k(v)$ and $\alpha_k(v)$ jointly. In particular, 
a spike-and-slab prior is assigned jointly on $(\alpha_k(v),\btheta_k(v)^T)^T$ as below,
\begin{align}\label{prior_u}
(\alpha_k(v),\btheta_k(v)^T)^T|\eta_{v,k}\sim \eta_{v,k} N(\bzero,\bL_k)+(1-\eta_{v,k})\delta_{\bzero},\:\:\eta_{v,k}\stackrel{i.i.d.}{\sim} Ber(\Delta_k)
\end{align}
where $\eta_{v,k}$ is the activation indicator shared by both $\alpha_k(v)$ and $\btheta_k(v)$, and 
$\bL_{k}$ is a covariance matrix of order $(R+1)\times (R+1)$, where $\bL_k=\left(\begin{array}{ll}
L_{k,11} & \bL_{k,12}\\
\bL_{k,12}^T & \bL_{k,22}
\end{array}\right)$. Here $L_{k,11}$ and $\bL_{k,22}$ are prior variances for $\alpha_k(v)$ and $\btheta_k(v)$ for the slab components, respectively, and $\bL_{k,12}$ is the prior covariance between $\alpha_k(v)$ and $\btheta_k(v)$. Notably, $\eta_{v,k}=0$ implies that the $k$th predictor of interest is unrelated to the node $\mathcal{N}(\bs_v)$, i.e., the node $\mathcal{N}(\bs_v)$ is uninfluential. The parameter $\Delta_{k}$ corresponds to the probability of the nonzero mixture component in (\ref{prior_u}). Apart from drawing inference on influential nodes using $\{\eta_{v,k}:v=1,...,V;\:k=1,...,p\}$, (\ref{prior_u}) is instrumental in the construction of predictor dependent association between the network and the nodal attribute. In fact, the predictor-dependent covariance between the attribute at node $\mathcal{N}(\bs_v)$ and network latent effect for node $\mathcal{N}(\bs_v)$ corresponding to all predictors is given by
a $R\times p$ dimensional matrix
\begin{align}\label{eq:covariance_network_nodal}
Cov\left(z_i(\bs_v), \left[\btheta_1(v):\cdots:\btheta_p(v)\right]\right)=
\left[x_{i,1}\Delta_1L_{1,12}^T:\cdots:x_{i,p}\Delta_pL_{p,12}^T\right],\:\:i=1,...,n.
\end{align}
Here, the $k$th column represents the correlation between the network latent effect and the nodal attribute at the $v$th node.

Previous work by \cite{fosdick2015testing} focuses on the unsupervised joint modeling of a single network and its nodal attributes. Their method simultaneously models node-specific latent effects and nodal attributes with a multivariate Gaussian distribution. In this formulation, the node-specific latent effects and nodal attributes for all nodes are inferred to be independent if and only if the cross-covariance matrix of the multivariate Gaussian distribution is zero. They conduct inference on the cross-covariance matrix using a likelihood ratio test. \cite{guhaniyogi2020joint}  and \cite{zhang2022joint} extend this approach to impose dependence between network and nodal attributes through shared latent effects.

While our proposed approach shares some similarities with the aforementioned literature in jointly modeling the network's node-specific latent effects $\btheta_k(v)$ and the coefficients $\alpha_k(v)$ for the nodal attribute, it differs in several fundamental aspects. The primary distinction lies in the nature of joint learning. In our framework, joint learning is predictor-dependent, focusing on the simultaneous estimation of coefficients for both the main predictors of interest and the auxiliary predictors.

Another key difference concerns the inferential objective. Our method is designed to identify network nodes associated with a  predictor, leveraging joint information from both the network structure and the nodal attribute. Our approach accounts for the dependence between the network and the nodal attribute but does not focus on hypothesis testing to assess their dependence.
In contrast, \cite{guhaniyogi2020joint} and \cite{zhang2022joint} primarily aim to detect independence between the network and the nodal attribute across all nodes simultaneously. Their methods are primarily applied to social networks,  where inferential frameworks typically do not consider the spatial locations of nodes. In contrast, our framework explicitly incorporates spatial associations between nodal attributes across nodes, as elaborated in the next section.

\noindent\underline{\textbf{Prior construction on $\delta_i(\cdot)$.}} To impose a spatially dependent correlation structure on the nodal attribute across nodes, we assign a Gaussian process prior with a correlation kernel $\kappa(\cdot,\cdot;\zeta)$ and spatial variance $\tau_z^2$, denoted by 
$\delta_i(\cdot)\sim GP(0,\tau_z^2\kappa(\cdot,\cdot;\zeta))$, independently over $i=1,...,n$. The prior construction implies that $Cov(z_i(\bs_u),z_i(\bs_v))=\tau_z^2\kappa(\bs_u,\bs_v;\zeta)$, where $\zeta$ is the length-scale parameter controlling the degree of spatial correlation between the nodal attribute at nodes $\mathcal{N}(\bs_u)$ and $\mathcal{N}(\bs_v)$. We employ the widely used exponential covariance kernel given by 
 $\kappa(\bs_u,\bs_v;\zeta)=\exp(-\zeta||\bs_u-\bs_v||)$. A discrete uniform prior is set for $\zeta$. 
The covariance matrix $\bL_k$ is assigned a IW($\nu,\bSigma_{\bL})$ prior. 
The prior specification is completed by assigning $\tau_y^2,\tau_z^2\sim IG(a,b)$ and placing flat priors on $\mu_y,\mu_z, \gamma_{1,y},...,\gamma_{q,y},\gamma_{1,z},...,\gamma_{q,z}$, i.e., $f(\mu_y),f(\mu_z), f(\gamma_{1,y}),...,f(\gamma_{q,y}),f(\gamma_{1,z}),...,f(\gamma_{q,z})\propto 1$.

\begin{figure}
    \centering
    \includegraphics[width=0.8\linewidth]{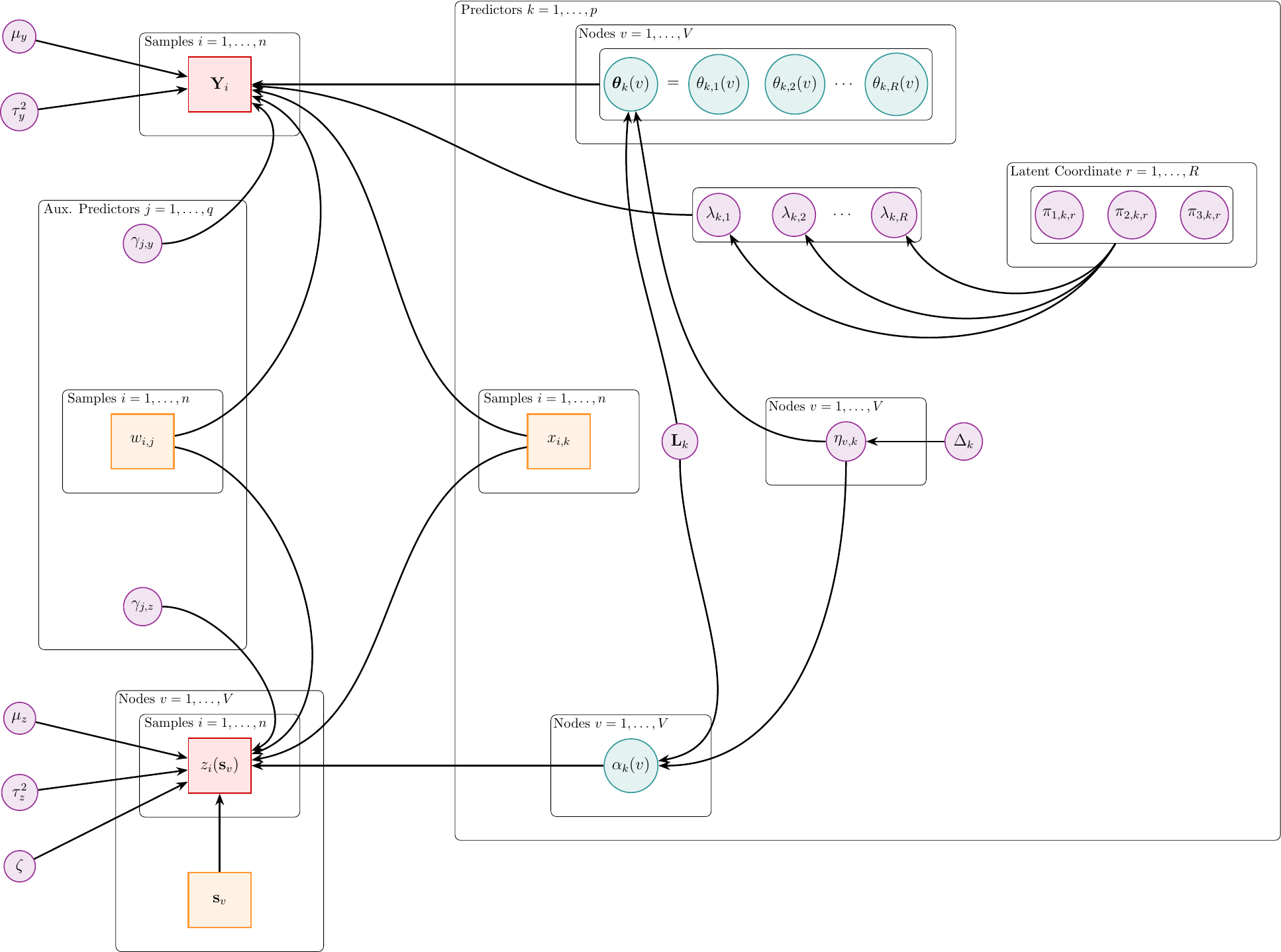}
    \caption{Diagram of the dependencies between observations (red), predictors (orange), latent variables (teal), and parameters (violet).}
    \label{model_diagram}
\end{figure}

\section{Posterior Computation}\label{sec:posterior_comp}

Although Sections~\ref{sec:model_formulation} and \ref{sec:prior_structure} present a model formulation and prior construction that do not permit a closed-form full posterior distribution of the parameters, all parameters have closed-form full conditional distributions. Consequently, the model computation is carried out using a Gibbs sampler. The full conditional distributions to perform Markov Chain Monte Carlo (MCMC) sampling are provided in the supplementary material.

Due to rapid convergence of the MCMC sampler, it is executed for a total of $500$ iterations, with the initial $200$ iterations discarded as burn-in, and the inference is based on the remaining $300$ post burn-in iterations. Across all simulation scenarios, an average effective sample size of $235$ is observed, indicating relatively uncorrelated post burn-in MCMC samples.

Our code is implemented exclusively using R without using any C++, Fortran, or Python interface. The computations are conducted on a cluster computing server containing two AMD EPYC 7763 processors with a combined 128 cores and 256 threads, along with 1 TB of RAM. Multiple replications of the model are executed in parallel architecture using the R packages \emph{doparallel} and \emph{foreach}. The average computation time for running 500 MCMC iterations with $V=20$ is 30 minutes across all simulations.

To provide the posterior probability of node $\mathcal{N}(\bs_v)$ being influentially associated with the $k$th predictor of interest, we calculate the empirically estimated $P(\eta_{v,k}=1|Data)$, expressed as $\frac{1}{F}\sum_{f=1}^F \eta_{v,k}^{(f)}$. Here, $\eta_{v,k}^{(1)},\ldots,\eta_{v,k}^{(F)}$ represent the $F$ post burn-in MCMC samples. Node $\mathcal{N}(\bs_v)$ is considered influentially related to the $k$th predictor if the empirically estimated posterior probability surpasses $0.5$, in accordance with the median probability rule for variable selection \citep{barbieri2004optimal}.  However, this cutoff may be revised should a more compelling, scientifically validated cutoff become available for the specific problem under consideration.
The posterior distributions of $\alpha_k(v)$ and $\beta_k(u,v)$  are empirically estimated from the post burn-in MCMC samples of $\{\alpha_k(v)^{(f)}:v=1,..,V\}$, $\{\btheta_k(v)^{(f)}:v=1,..,V\}$ and $\{\lambda_{k,r}^{(f)}:r=1,..,R\}$. 
Additionally, we provide an empirical estimation of the spatial correlation between the nodal attribute at different nodes. For this purpose, from post burn-in samples $\zeta^{(f)},\tau_z^{(f)2},\mu_z^{(f)},\{\alpha_k(v)^{(f)}:k=1,\ldots,p\},\{\gamma_{j,z}^{(f)}:j=1,\ldots,q\}$, $f=1,\ldots,F$, we draw posterior predictive samples $(z_i(\bs_v)^{(f)},\ldots,z_i(\bs_v)^{(f)})^T$ for $i=1,\ldots,n$. For any pair of locations $\bs_v$ and $\bs_{v'}$, $v,v'=1,..,V$, an empirical correlation coefficient is then computed to estimate the decay of spatial correlation over spatial distance $||\bs_v-\bs_{v'}||$.

\section{Simulation Studies}

In this section, we evaluate the performance of our proposed method and compare it with several competing approaches across a range of simulation scenarios. 
This will allow us to assess the method’s accuracy in identifying influential nodes, its ability to estimate predictor coefficients corresponding to both network and nodal attributes, and its effectiveness in capturing spatial correlations between attributes across different network nodes.
It will also highlight the importance of accounting for the network structure, joint modeling of network and nodal attributes, and spatial correlations of attributes over nodes. 

\noindent\underline{\textbf{Simulated Data Generation.}} 
The simulation study sets the number of predictors of interest $p=1$ and the number of auxiliary predictors $q=2$. For the $i$th sample, the predictor of interest $x_i$, and the auxiliary predictors $\bw_i$ are drawn from $N(0,1)$ and $N(\textbf{0},\bI_{q})$, respectively. 

Let $\Delta^*$ be the true probability of node inclusion, with the quantity $(1 - \Delta^*)$ being referred to as the \emph{node sparsity parameter}. The node inclusion indicators, $\eta^{*}_{1},...,\eta^{*}_{V}$, are drawn from $Ber(\Delta^*)$. The true node-specific latent network effect and the true predictor coefficient for nodal attributes are drawn together
$\begin{bmatrix}
     \alpha^{*}(v) \\
      \btheta^{*}(v)
  \end{bmatrix} $  from $\eta^{*}_{v} N(\textbf{0}, \bL^*)
  + (1 - \eta^{*}_{v})\delta_{\bzero}$, , for $v=1,...,V$,  where $R^{*}$ is the true latent dimension and $\bL^*$ is sampled from $IW(R^* + 2, \bI_{R^{*}+1})$. For all simulations, we set $R^{*} = 4$. The network coefficient corresponding to the edge connecting the $u$th and $v$th nodes, $\beta^*(u,v)$, is constructed from the $u$th and $v$th node-specific latent effects as $\beta^*(u,v) = {\btheta^{*}(u)}^{T}\btheta^{*}(v)$, for $1 \leq u < v\leq V$. We also set  $\beta^*(u,v) = \beta^*(v,u)$  and $\beta^*(v,v) = 0$ to satisfy symmetry with zero diagonal entries for the network coefficient matrix. 

  For $i=1,..,n$, we simulate the network edge $y_i(u,v)$, following equation
  (\ref{network_equation}) with $p=1$, $q=2$ and $\beta(u,v)$, $\gamma_{j,y}$, $\mu_y$ and $\tau_y^2$ replaced by their true values $\beta^*(u,v)$, $\gamma_{j,y}^*$, $\mu_y^*$ and $\tau_y^{2*}$, respectively. 
To simulate the nodal attributes, three-dimensional spatial coordinates for the $V$ nodes, $\bs^*_1,...,\bs^*_V$, are first sampled from $N(\bzero, \bI)$. Then, the true correlated spatial effects $\bdelta_i^*=(\delta_i^*(\bs_1),...,\delta_i^*(\bs_V))^T$ are simulated from $N(\bzero,
  \tau_z^{2*}\bSigma^*)$, where $\bSigma^*$ is a $V\times V$ matrix  with the $(u,v)$th entry given by $\exp(-\zeta^*||\bs_u-\bs_v||)$. Here, $\zeta^*$ is the scale parameter that determines the range of spatial dependence among nodal attributes across the nodes in the simulated data. Subsequently, for $i=1,..,n$, the nodal attribute $z_i(\bs_v)$ is simulated following (\ref{structural_equation}) with $\alpha(v)$, $\gamma_{j,z}$, $\mu_z$ and $\delta_i(\bs_v)$ replaced by their true values $\alpha^*(v)$, $\gamma_{j,z}^*$, $\mu_z^*$ and $\delta_i^*(\bs_v)$, respectively.
  

\noindent \underline{\textbf{Simulation Scenarios.}}\label{sec:simulations}
Our extensive simulation studies reveal that the performance of our approach is influenced by the true node sparsity and scale parameter controlling the range of spatial dependence. Hence, we use various combinations of node sparsity $(1-\Delta^*)$ and the scale parameter $\zeta^*$, as specified in Table~\ref{Tab:sim_scenario}. In all simulations, we set $\tau^{2*}_y = 1$, $\tau^{2*}_z = 9$, $\bgamma_y^*=(\gamma_{1,y}^*, \gamma_{2,y}^*)^T = (0.2, 0.5)^T$ and $\bgamma_z^* = (\gamma_{1,z}^*, \gamma_{2,z}^*)^T = (0.1, 0.4)^T$, $\mu_y^*=\mu_z^*=0$.   

\begin{table}[H]
\centering
\begin{tabular}{|c|c|c|c|c|c|c|c|}
\hline
Scenario & 1 & 2 & 3 & 4 & 5 & 6 & 7 \\
\hline
$1-\Delta^*$ & 0.8 & 0.7 & 0.7 & 0.5 & 0.5 & 0.4 & 0.3 \\
\hline
$\zeta^*$ & 0.05 & 0.1 & 0.2 & 0.1 & 0.2 & 0.05 & 0.05 \\
\hline
\end{tabular}
\caption{Simulation scenarios varying the node sparsity $(1 - \Delta^{*})$ and spatial scale parameter $\zeta^*$.  }\label{Tab:sim_scenario}
\end{table}

\noindent \underline{\textbf{Competitors.}} 
A comparison of our proposed method, referred to as the {\emph{spatial joint model}}, against several competing methods reveals the critical importance of its key features: its ability to account for network structure, jointly model network and nodal attributes, and capture spatial correlations of attributes across nodes. 
The first competitor, referred to as the \emph{independent network model}, involves separately fitting models (\ref{network_equation}) and (\ref{structural_equation}). Comparison with the independent network model will allow for assessment of the impact of jointly estimating these models. The second competitor, referred to as the \emph{independent tensor model}, similarly involves independent modeling of each object, but instead of fitting (\ref{network_equation}), it treats the network outcome as a $V \times V$ tensor, and fits a tensor response regression  using the \texttt{TRES} package in R \citep{zeng2021TRES}. 
We employ (\ref{structural_equation}) for modeling nodal attributes for this competitor. This competitor will assess both the efficacy of joint modeling and the impact of violating the symmetry assumption on the network outcome. The next competitor, referred to as the \emph{non-spatial joint model}, fits our joint model but assumes no spatial association between nodal attributes over nodes. This will assess the utility of accounting for spatial correlation of nodal attributes in modeling.        

\noindent \underline{\textbf{Metrics of Comparison.}}
We estimate the probability of a node being influential, $P(\eta_v=1|Data)$, to assess the accuracy of node selection for our model and the competitors. 
To assess the efficacy of our method versus the competitors in coefficient estimation and uncertainty quantification, we calculate the scaled mean squared error (MSE) of the estimated coefficients for the network and node-specific components of the model and acquire the coverage and length of 95\% credible intervals for these quantities. The MSE will be calculated as $\dfrac{||\widehat{\bbeta} - \bbeta^{*}||^{2}}{||\bbeta^{*}||^{2}}$ for the network model coefficients and as $\dfrac{||\widehat{\balpha} - \balpha^{*}||^{2}}{||\balpha^{*}||^{2}}$ for the node-specific model coefficients, where $\widehat{\bbeta}$ and $\widehat{\balpha}$ represent point estimates for $\bbeta$ and $\balpha$, respectively. For the Bayesian models, these point estimates are the posterior means of the quantities, while for the tensor learning method, the network coefficients are calculated through a frequentist point estimate.  We obtain the length and coverage of posterior 95\% credible intervals averaged over edge coefficients from the post burn-in MCMC samples. For the frequentist competitors, we compute the coverage and length of the confidence intervals with an asymptotic coverage of 95\%.

\subsection{Simulation Results}
\subsubsection{Node Selection}
Table \ref{table-node-selection} shows the results of node selection across the seven different simulation combinations for the spatial joint model, the non-spatial joint model, and the independent network model. The entries in the table contain the estimated posterior probabilities $P(\eta_{v}=1|Data)$ for $v=1,...,20$. The independent tensor model is not designed to  offer node selection probabilities.

The results demonstrate that, across the seven combinations, the spatial joint model accurately determines whether a node is influential or non-influential in each case, using a threshold of 0.5 to assess significance. The only exception is a false negative in Scenario 6, which was also misidentified by the other models. 
In all other cases, the posterior probabilities of truly influential nodes are high, with those of non-influential nodes being smaller and in many cases close to 0. There is a little more uncertainty in Scenarios 1 and 2, which correspond to high levels of node sparsity. The non-spatial joint model demonstrates competitive performance with the spatial joint model except in Scenarios 2, 4, and 6, where it has one false positive, one false positive, and two false negatives respectively. Under these scenarios, the non-spatial model makes errors due to not accounting for the strong spatial correlation among nodal attributes. The independent network model incurs two additional misclassifications compared to the non-spatial joint model in Scenario 1, and generally incurs little higher uncertainties (i.e., little higher $P(\eta_v=1|Data))$ for un-influential nodes. This highlights the benefits of incorporating further network information and performing joint modeling of the data for node selection, particularly when the level of node sparsity in the network is higher.

\begin{table}[H]
\subfloat[Spatial Joint Model] 
{\resizebox{0.49\linewidth}{!}{%
\setlength{\tabcolsep}{3pt}
\begin{tabular}{c|c|c|c|c|c|c|c}
    Node & \makecell{Scen. 1 \\ $1-\Delta^* = 0.8$ \\ $\zeta^*= 0.05$} & \makecell{Scen. 2 \\ $1-\Delta^* = 0.7$ \\ $\zeta^*= 0.1$} & \makecell{Scen. 3 \\ $1-\Delta^* = 0.7$ \\ $\zeta^*= 0.2$} & \makecell{Scen. 4 \\ $1-\Delta^* = 0.5$ \\ $\zeta^*= 0.1$} & \makecell{Scen. 5 \\ $1-\Delta^* = 0.5$ \\ $\zeta^*= 0.2$} & \makecell{Scen. 6 \\ $1-\Delta^* = 0.4$ \\ $\zeta^*= 0.05$} & \makecell{Scen. 7 \\ $1-\Delta^* = 0.3$ \\ $\zeta^*= 0.05$} \\
        \hline
  1 &  0.0588 &  0.0196 &  0.0000 & \cellcolor{red!25} 1.0000 & \cellcolor{red!25} 1.0000 & \cellcolor{red!25} 1.0000  & \cellcolor{red!25} 1.0000 \\
      \hline
  2 &  0.0980 & \cellcolor{red!25} 1.0000 &  0.1176 &  0.0784 & \cellcolor{red!25} 0.6863 & \cellcolor{red!25} 1.0000  &  \cellcolor{red!25} 1.0000 \\
      \hline
  3 &   0.0196 &  0.2353 &  0.0000 & \cellcolor{red!25} 1.0000 &  0.0588 & \cellcolor{red!25} 1.0000    &  \cellcolor{red!25} 1.0000 \\
      \hline
  4 &  0.0784 &  0.0000 & \cellcolor{red!25} 1.0000 &  0.0196 & \cellcolor{red!25} 1.0000 &  0.0000  & \cellcolor{red!25}  1.0000 \\
      \hline
 5 &  0.0196 &  0.0196 & \cellcolor{red!25} 1.0000 &  0.0000 & \cellcolor{red!25} 1.0000 & \cellcolor{red!25} 1.0000   & \cellcolor{red!25} 1.0000 \\
     \hline
 6 & \cellcolor{red!25} 1.0000 & \cellcolor{red!25} 1.0000 & \cellcolor{red!25} 1.0000 &  0.0000 & \cellcolor{red!25} 1.0000 &  0.0000   &  0.0000 \\
     \hline
 7 & 0.0000 & \cellcolor{red!25} 1.0000 & 0.0000 & \cellcolor{red!25} 1.0000 & 0.0196 & \cellcolor{red!25} 1.0000    & \cellcolor{red!25} 1.0000 \\
     \hline
 8 &  0.0000 &  0.0000 & \cellcolor{red!25} 1.0000 &  0.1373 &  0.0980 & \cellcolor{red!25} 1.0000   & \cellcolor{red!25} 1.0000 \\
     \hline
 9 &  0.2157 &  0.0588 & \cellcolor{red!25} 1.0000 &  0.0000 &  0.0392 & \cellcolor{red!25} 0.9412   & \cellcolor{red!25} 1.0000 \\
     \hline
10 &  0.0000 & \cellcolor{red!25} 1.0000 & \cellcolor{red!25} 1.0000 & \cellcolor{red!25} 1.0000 &  0.0588 &  0.0000  &   0.0000 \\
    \hline
11 &  0.0392 &  0.2353 &  0.0000 &  0.0392 &  0.2941 &  0.0196  &   0.0000 \\
    \hline
12 & \cellcolor{red!25} 1.0000 &  0.0588 &  0.0000 &  \cellcolor{red!25} 1.0000 & \cellcolor{red!25} 1.0000 & \cellcolor{red!25} 1.0000    & \cellcolor{red!25} 1.0000 \\
    \hline
13 &   0.4314 & \cellcolor{red!25} 1.0000 &  0.0000 &  \cellcolor{red!25} 1.0000 &  0.0000 &  0.0196 &  0.0000 \\
    \hline
14 & \cellcolor{red!25} 1.0000 & \cellcolor{red!25} 1.0000 &  0.0000 & \cellcolor{red!25} 1.0000 &  0.0392 & \cellcolor{red!25} 1.0000   & \cellcolor{red!25} 1.0000 \\
    \hline
15 &  0.0196 &  0.0000 &  0.0196 & \cellcolor{red!25} 1.0000 & \cellcolor{red!25} 1.0000 & \cellcolor{red!25} 0.0980  & \cellcolor{red!25} 1.0000 \\
    \hline
16 &  0.0588 &  0.0000 &  0.0000 &  0.0196 & \cellcolor{red!25} 1.0000 & \cellcolor{red!25} 1.0000  & \cellcolor{red!25} 1.0000 \\
    \hline
17 &  0.0000 &  0.0000 &  0.0588 & \cellcolor{red!25} 1.0000 & \cellcolor{red!25} 1.0000 &  0.0392 &  0.0000 \\
    \hline
18 &  0.0392 &  0.0196 &  0.0000 & \cellcolor{red!25} 1.0000 &  0.0588 &  0.0392  &  0.0000 \\
    \hline
19 &  0.0196 &  0.0196 &  0.0000 &  0.4314 &  0.1176 &  0.0000  &  \cellcolor{red!25} 1.0000 \\
    \hline
20 & \cellcolor{red!25} 0.5882 &  0.0588 &  0.0000 &  0.0588 & \cellcolor{red!25} 0.9020 & \cellcolor{red!25} 1.0000 &  \cellcolor{red!25} 1.0000 
\end{tabular}}}
\hfill
\subfloat[Non-spatial Joint Model]
{\resizebox{0.49\linewidth}{!}{%
\setlength{\tabcolsep}{3pt}
\begin{tabular}{c|c|c|c|c|c|c|c}
Node & \makecell{Scen. 1 \\ $1-\Delta^* = 0.8$ \\ $\zeta^*= 0.05$} & \makecell{Scen. 2 \\ $1-\Delta^* = 0.7$ \\ $\zeta^*= 0.1$} & \makecell{Scen. 3 \\ $1-\Delta^* = 0.7$ \\ $\zeta^*= 0.2$} & \makecell{Scen. 4 \\ $1-\Delta^* = 0.5$ \\ $\zeta^*= 0.1$} & \makecell{Scen. 5 \\ $1-\Delta^* = 0.5$ \\ $\zeta^*= 0.2$} & \makecell{Scen. 6 \\ $1-\Delta^* = 0.4$ \\ $\zeta^*= 0.05$} & \makecell{Scen. 7 \\ $1-\Delta^* = 0.3$ \\ $\zeta^*= 0.05$} \\
    \hline
 1 & 0.0652 & 0.1957 & 0.0000 & \cellcolor{red!25} 1.0000 & \cellcolor{red!25} 1.0000  & \cellcolor{red!25} 1.0000   & \cellcolor{red!25} 1.0000 \\
    \hline
 2 & 0.1087 & \cellcolor{red!25} 1.0000 & 0.0435 & 0.0435 & \cellcolor{red!25} 0.7174 & \cellcolor{red!25} 1.0000   & \cellcolor{red!25} 1.0000 \\
    \hline
 3 & 0.0217 & 0.6087 & 0.0000 & \cellcolor{red!25} 1.0000 & 0.0217 & \cellcolor{red!25} 1.0000   & \cellcolor{red!25} 1.0000 \\
    \hline
 4 & 0.0435 & 0.0217 & \cellcolor{red!25} 1.0000 & 0.0435 & \cellcolor{red!25} 1.0000 & 0.0000   &  \cellcolor{red!25} 1.0000 \\
    \hline
 5 & 0.0000 & 0.0217 & \cellcolor{red!25} 1.0000 & 0.0000 & \cellcolor{red!25} 1.0000 & \cellcolor{red!25} 1.0000   & \cellcolor{red!25} 1.0000 \\
    \hline
 6 & \cellcolor{red!25} 1.0000 &  \cellcolor{red!25}1.0000 & \cellcolor{red!25} 1.0000 & 0.0217 & \cellcolor{red!25} 1.0000 & 0.0000   &  0.0000 \\
    \hline
 7 & 0.0000 & \cellcolor{red!25} 1.0000 & 0.0000 &  \cellcolor{red!25} 1.0000 & 0.0000 & \cellcolor{red!25} 1.0000   & \cellcolor{red!25} 1.0000 \\
    \hline
 8 & 0.0000 & 0.1739 & \cellcolor{red!25} 1.0000 & 0.1739 & 0.1087 & \cellcolor{red!25} 1.0000  &   \cellcolor{red!25} 1.0000 \\
    \hline
 9 & 0.1522 & 0.2174 & \cellcolor{red!25} 1.0000 & 0.0217 & 0.0652 & \cellcolor{red!25} 0.3913  &   \cellcolor{red!25} 1.0000 \\
    \hline
10 & 0.1522 & \cellcolor{red!25} 1.0000 & \cellcolor{red!25} 1.0000 & \cellcolor{red!25} 1.0000 & 0.0217 & 0.0000   &  0.0000 \\
   \hline
11 & 0.0217 & 0.3478 & 0.0000 & 0.0000 & 0.3261 &  0.0000   &  0.0000 \\
   \hline
12 & \cellcolor{red!25} 1.0000 & 0.0000 & 0.0000 & \cellcolor{red!25} 1.0000 & \cellcolor{red!25} 1.0000 & \cellcolor{red!25} 1.0000   & \cellcolor{red!25} 1.0000 \\
   \hline
13 & 0.4565 & \cellcolor{red!25} 1.0000 & 0.0000 & \cellcolor{red!25} 1.0000 & 0.0217 & 0.0000   &  0.0000 \\
   \hline
14 & \cellcolor{red!25} 1.0000 & \cellcolor{red!25} 1.0000 & 0.0000 & \cellcolor{red!25} 1.0000 & 0.0217 & 
 \cellcolor{red!25} 1.0000  &  \cellcolor{red!25} 1.0000 \\
   \hline
15 &  0.0217 & 0.0652 & 0.0000 & \cellcolor{red!25} 1.0000 & \cellcolor{red!25} 1.0000 & \cellcolor{red!25} 0.0652  &  \cellcolor{red!25} 1.0000 \\
   \hline
16 & 0.0652 & 0.0652 & 0.0000 & 0.0435 & \cellcolor{red!25} 1.0000 & \cellcolor{red!25} 1.0000   & \cellcolor{red!25} 1.0000 \\
   \hline
17 & 0.0217 & 0.0217 & 0.0217 & \cellcolor{red!25} 1.0000 & \cellcolor{red!25} 1.0000 & 0.0000  &  0.0000 \\
   \hline
18 & 0.0217 & 0.0652 & 0.0000 & \cellcolor{red!25} 1.0000 & 0.0435 & 0.0000   & 0.0000 \\
   \hline
19 & 0.0217 & 0.0870 & 0.0217 & 0.5435 & 0.2174 & 0.0000   & \cellcolor{red!25} 1.0000 \\ 
   \hline
20 & \cellcolor{red!25} 0.5870 & 0.0652 & 0.0000 & 0.0435 & \cellcolor{red!25} 0.9783 & \cellcolor{red!25} 1.0000  &  \cellcolor{red!25} 1.0000  
\end{tabular}}}
\hfill
\centering
\subfloat[Independent Network Model]
{\resizebox{0.49\linewidth}{!}{%
\setlength{\tabcolsep}{3pt}
\begin{tabular}{c|c|c|c|c|c|c|c}
Node & \makecell{Scen. 1 \\ $1-\Delta^* = 0.8$ \\ $\zeta^*= 0.05$} & \makecell{Scen. 2 \\ $1-\Delta^* = 0.7$ \\ $\zeta^*= 0.1$} & \makecell{Scen. 3 \\ $1-\Delta^* = 0.7$ \\ $\zeta^*= 0.2$} & \makecell{Scen. 4 \\ $1-\Delta^* = 0.5$ \\ $\zeta^*= 0.1$} & \makecell{Scen. 5 \\ $1-\Delta^* = 0.5$ \\ $\zeta^*= 0.2$} & \makecell{Scen. 6 \\ $1-\Delta^* = 0.4$ \\ $\zeta^*= 0.05$} & \makecell{Scen. 7 \\ $1-\Delta^* = 0.3$ \\ $\zeta^*= 0.05$} \\
    \hline  
    1 & 0.0732  & 0.0000  & 0.0000  & \cellcolor{red!25} 1.0000  & \cellcolor{red!25} 1.0000  & \cellcolor{red!25} 1.0000   &  \cellcolor{red!25} 1.0000 \\
    \hline  
 2  & 0.5610 & \cellcolor{red!25} 1.0000 &  0.0488 &  0.2439 & \cellcolor{red!25} 0.6098 & \cellcolor{red!25} 1.0000  &  \cellcolor{red!25}  1.0000 \\
 \hline  
 3  & 0.0732 &  0.3659 &  0.0000  & \cellcolor{red!25} 1.0000  & 0.0732 & \cellcolor{red!25} 1.0000   & \cellcolor{red!25}  1.0000 \\
 \hline  
 4  & 0.4634  & 0.0000  & \cellcolor{red!25} 1.0000  & 0.0976  & \cellcolor{red!25} 1.0000  & 0.0000   & \cellcolor{red!25}  1.0000 \\
 \hline  
 5 &  0.0732  & 0.0000  & \cellcolor{red!25} 1.0000 &  0.0244  & \cellcolor{red!25} 1.0000  & \cellcolor{red!25} 1.0000   &  \cellcolor{red!25} 1.0000 \\
 \hline  
 6 &  \cellcolor{red!25} 1.0000  & \cellcolor{red!25} 1.0000 & \cellcolor{red!25} 1.0000 &  0.0244 & \cellcolor{red!25} 1.0000 &  0.0244   &   0.0000 \\
 \hline  
 7 &  0.0244 & \cellcolor{red!25} 1.0000  & 0.0000 &  \cellcolor{red!25} 1.0000 &  0.0244 & \cellcolor{red!25} 1.0000   &  \cellcolor{red!25} 1.0000 \\
 \hline  
 8  & 0.0732  & 0.0244  & \cellcolor{red!25} 1.0000  & 0.2439  & 0.0976  &  \cellcolor{red!25} 1.0000    &  \cellcolor{red!25} 1.0000 \\
 \hline  
 9  & 0.1951 &  0.0244  & \cellcolor{red!25} 1.0000  & 0.0244  & 0.0488 & \cellcolor{red!25} 1.0000   &  \cellcolor{red!25} 1.0000 \\
 \hline  
10  & 0.1951  & \cellcolor{red!25} 1.0000  & \cellcolor{red!25} 1.0000  &  \cellcolor{red!25} 1.0000  & 0.0488  & 0.0244  &  0.0000 \\
\hline  
11  & 0.0732  & 0.1951  & 0.0000  & 0.0488  & 0.1707  & 0.0000    &  0.0000 \\
\hline  
12 & \cellcolor{red!25} 1.0000 & 0.0244 & 0.0000 & \cellcolor{red!25} 1.0000 & \cellcolor{red!25} 1.0000 & \cellcolor{red!25} 1.0000   &  \cellcolor{red!25} 1.0000 \\
\hline  
13 & 0.3659 & \cellcolor{red!25} 1.0000 & 0.0000 & \cellcolor{red!25} 1.0000 & 0.0000 & 0.0000   &  0.0000 \\
\hline  
14 & \cellcolor{red!25} 1.0000 & \cellcolor{red!25} 1.0000 & 0.0000 & \cellcolor{red!25} 1.0000 & 0.0488 & \cellcolor{red!25} 1.0000  &  \cellcolor{red!25} 1.0000 \\
\hline  
15 & 0.0244 & 0.0000 & 0.0244 & \cellcolor{red!25} 0.9268 &\cellcolor{red!25} 0.9756 & \cellcolor{red!25} 0.1463  & \cellcolor{red!25} 1.0000 \\
\hline  
16 & 0.1951 & 0.0000 & 0.0244 & 0.0488 & \cellcolor{red!25} 1.0000 & \cellcolor{red!25} 1.0000 &  \cellcolor{red!25} 1.0000 \\
\hline  
17 & 0.0244 & 0.0244 & 0.1220 & \cellcolor{red!25} 1.0000 & \cellcolor{red!25} 1.0000 & 0.0000  &   0.0000 \\
\hline  
18 & 0.1220 & 0.0000 & 0.0000 & \cellcolor{red!25} 1.0000 & 0.0976 & 0.0000  &   0.0000 \\
\hline  
19 & 0.0488 & 0.0244 & 0.0244 & 0.4878 & 0.0976 & 0.0244  & \cellcolor{red!25} 1.0000 \\
\hline  
20 & \cellcolor{red!25} 0.4146 & 0.0000 & 0.0244 & 0.0244 & \cellcolor{red!25} 0.8049 & \cellcolor{red!25} 1.0000 & \cellcolor{red!25} 1.0000 
\end{tabular}}}
\caption{
Node selection results for $V = 20$ nodes under the seven simulation scenarios. We show the posterior probability of each node being active for spatial joint model, non-spatial joint model and independent network model. The colored cells correspond to the truly influential nodes. The table shows excellent node identification by the spatial joint model and competitive performance by non-spatial joint model except in Scenarios 2, 4, and 6, and the independent network model except in Scenario 1. The node selection results are not available for independent tensor learning by design.}
\label{table-node-selection}
\end{table}

\subsubsection{Estimation of Regression Coefficients}
\begin{figure}[H]
\centering
\subfloat[Truth (Scenario 1)]{
\includegraphics[width=0.32\linewidth]{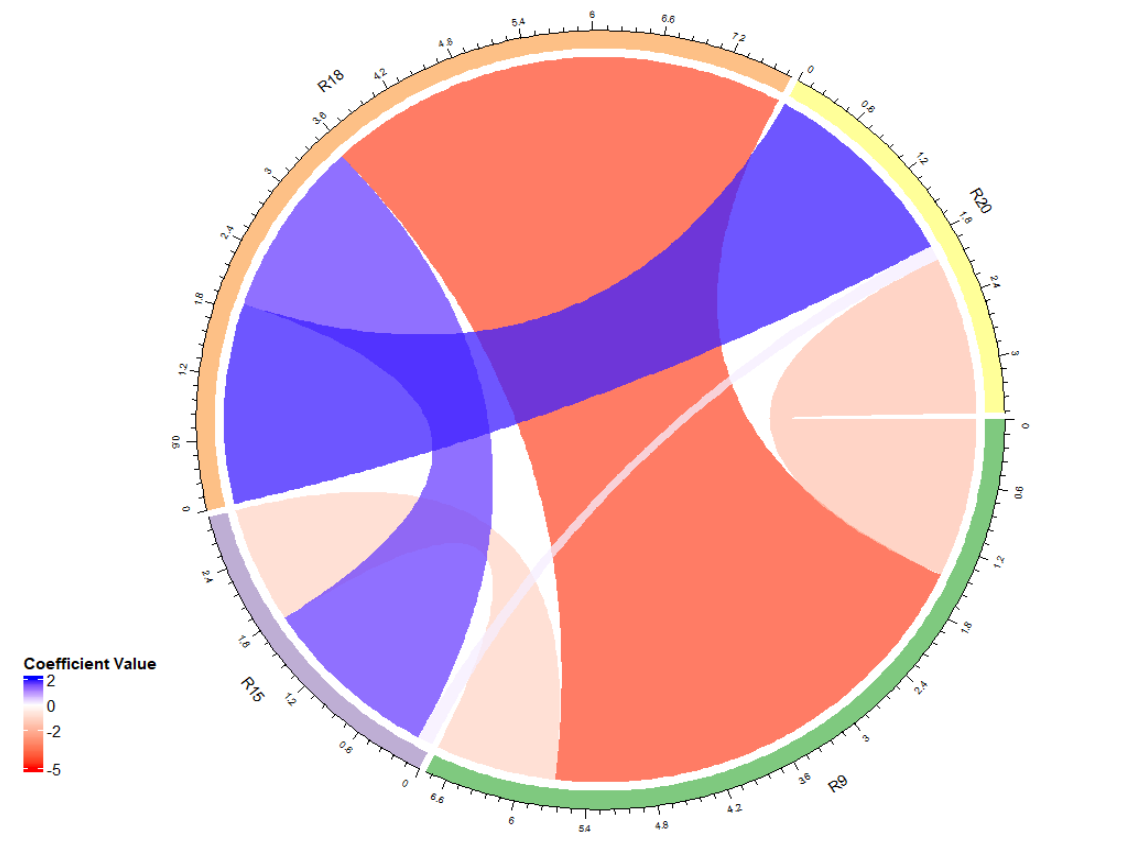}}
\subfloat[Truth (Scenario 4)]{
\includegraphics[width=0.32\linewidth]{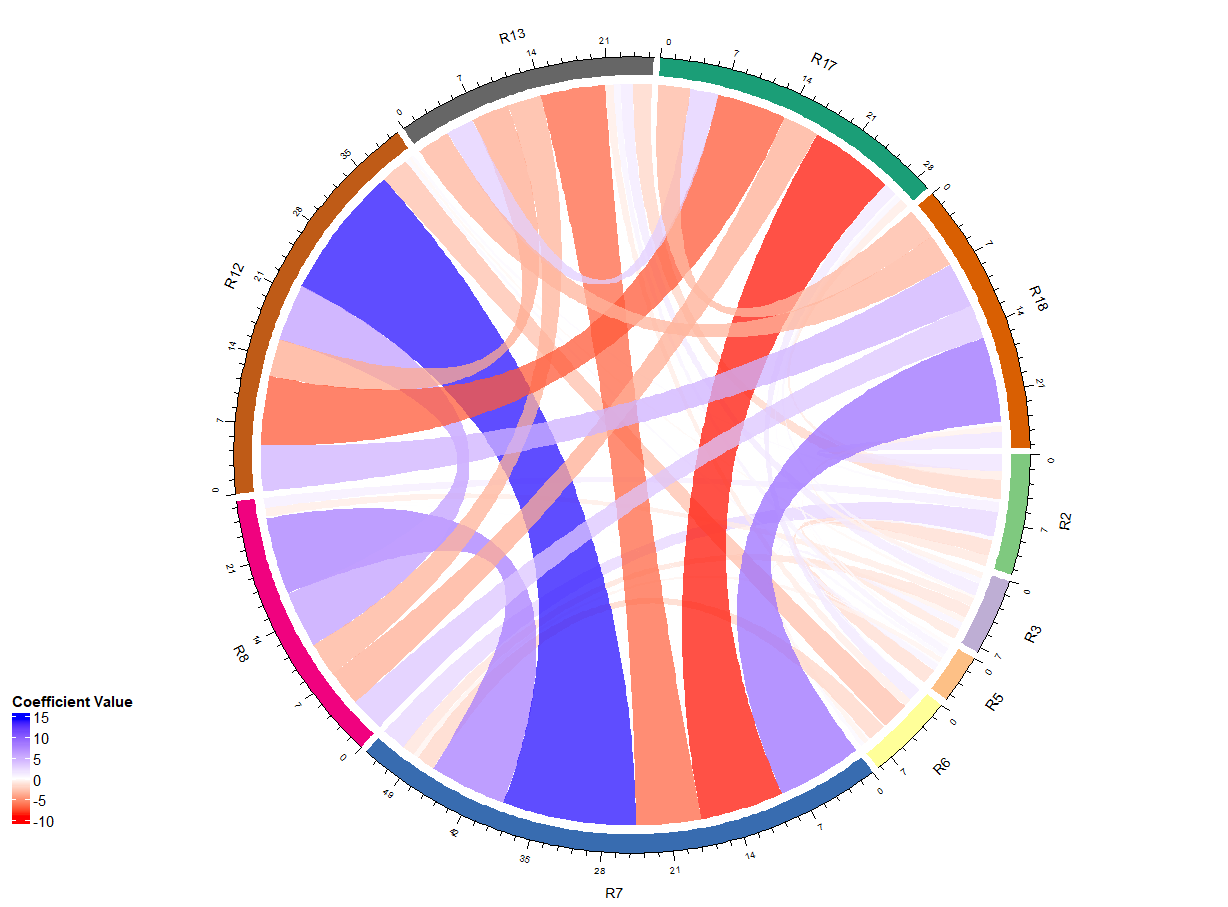}}
\subfloat[Truth (Scenario 7)]{
\includegraphics[width=0.32\linewidth]{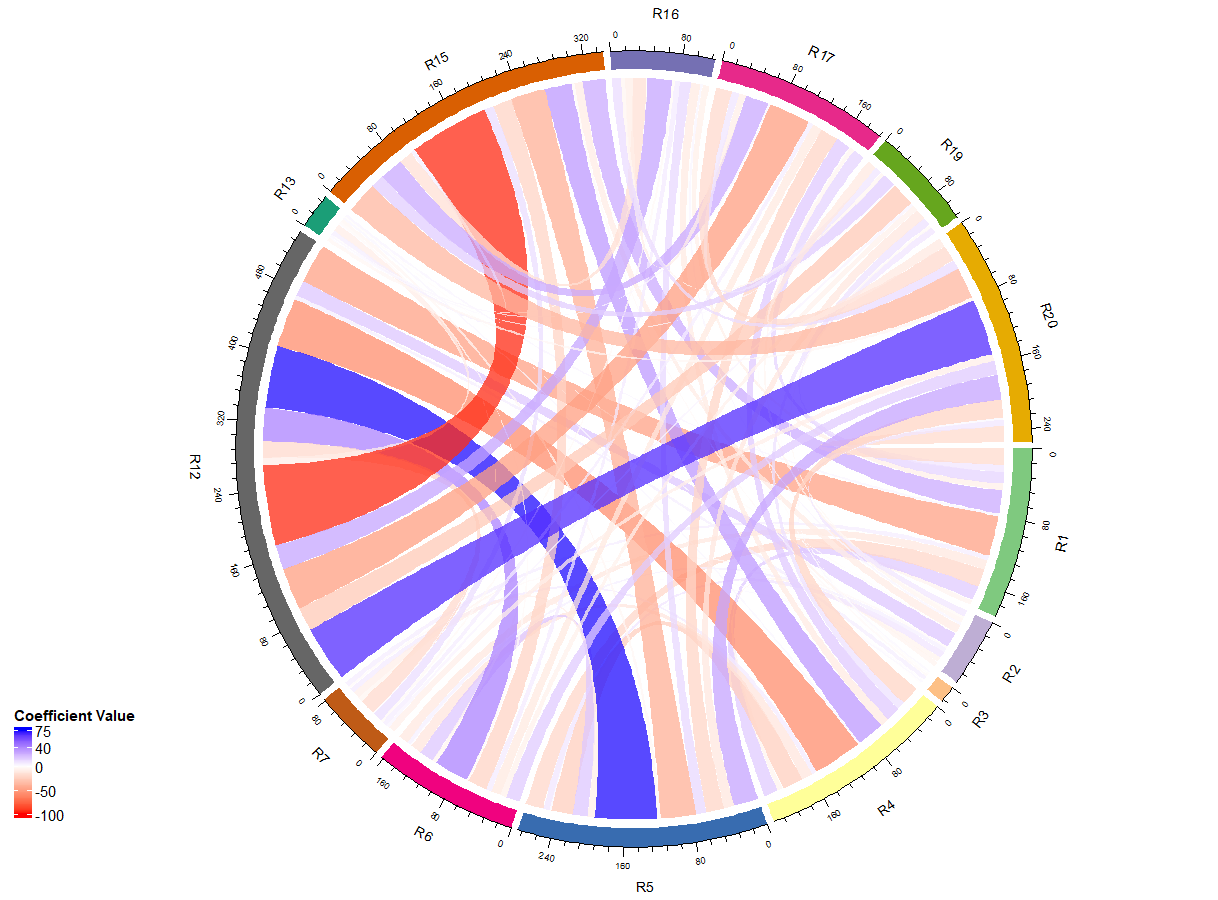}}
\hfill
\subfloat[Estimated (Scenario 1)]{
\includegraphics[width=0.32\linewidth]{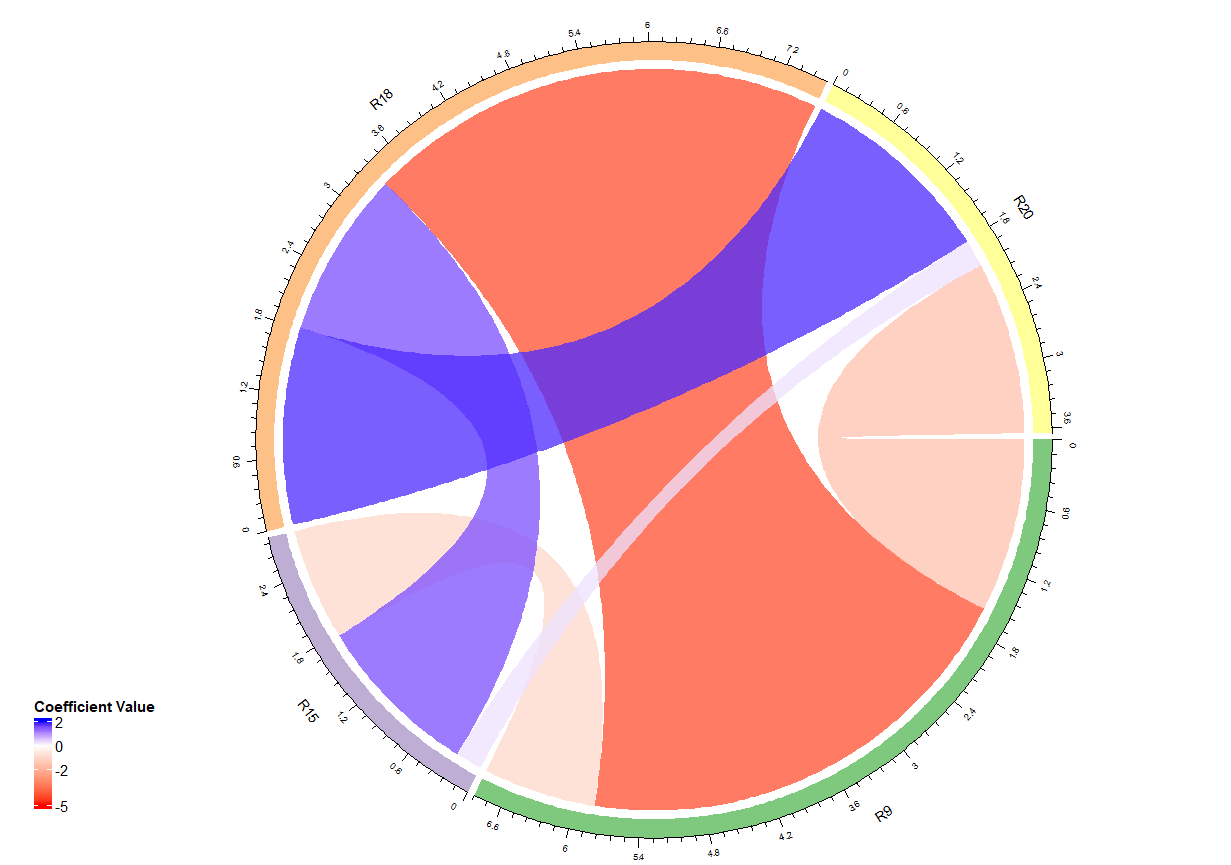}}
\subfloat[Estimated  (Scenario 4)]{
\includegraphics[width=0.32\linewidth]{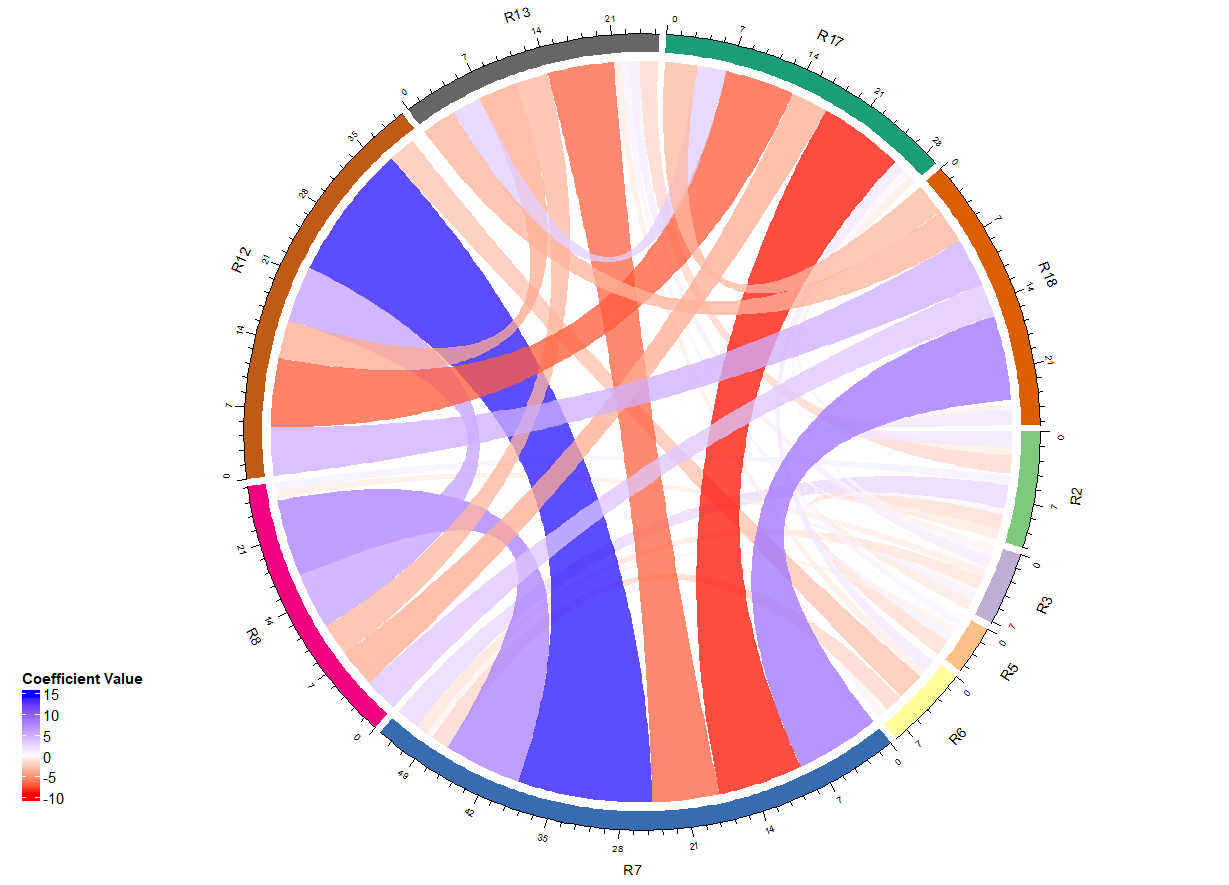}}
\subfloat[Estimated (Scenario 7)]{
\includegraphics[width=0.32\linewidth]{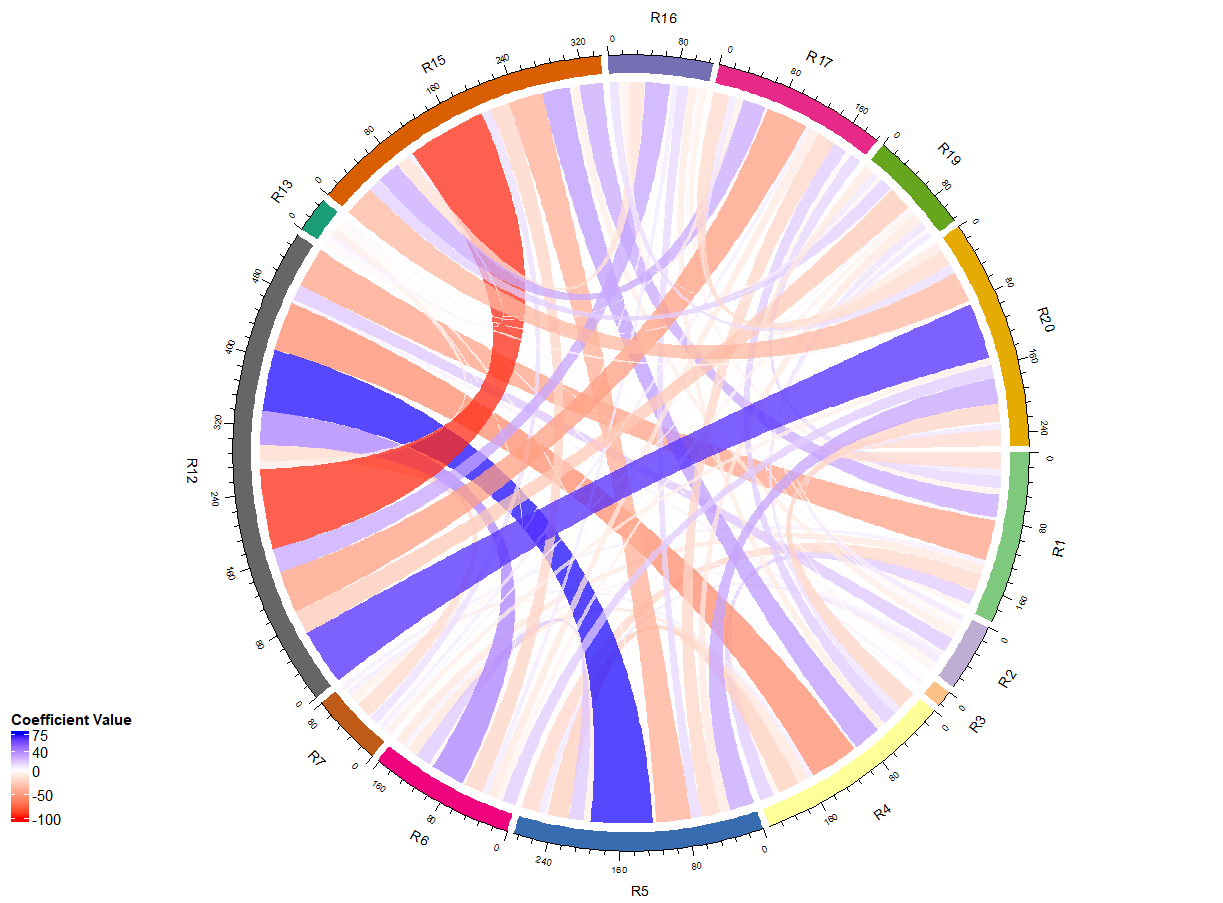}}
\caption{Circos plots illustrating the true and estimated covariate-dependent associations between influential network nodes, as captured by the network coefficients under Scenario 1 ($1-\Delta^* = 0.8$, $\zeta^*= 0.05$), Scenario 4 ($1-\Delta^* = 0.5$, $\zeta^*= 0.1$), and Scenario 7 ($1-\Delta^* = 0.3$, $\zeta^*= 0.05$). 
The top row displays the true network coefficients between influential network nodes, while the bottom row shows the corresponding estimates obtained from the proposed model. Across all scenarios, the model demonstrates highly accurate recovery of the underlying associations.}
\label{sim-circos-fig}
\end{figure}

Figure~\ref{sim-circos-fig} presents circos plots comparing the true and estimated association strengths $\beta(u,v)$ between influential nodes across selected simulation scenarios. For the sake of brevity, results are shown only for Scenarios $1, 4$, and $7$, which represent a broad spectrum of node sparsity levels, corresponding to true node sparsity values of $0.8, 0.5$, and $0.3$, respectively. In all cases, the estimated associations closely mirror the true values, highlighting the method’s strong ability to recover node relationships across varying degrees of sparsity.

\begin{figure}[h!]
\centering
\subfloat[Scaled MSE of estimating the entries of $\bbeta$]{%
    \includegraphics[width=0.49\linewidth]{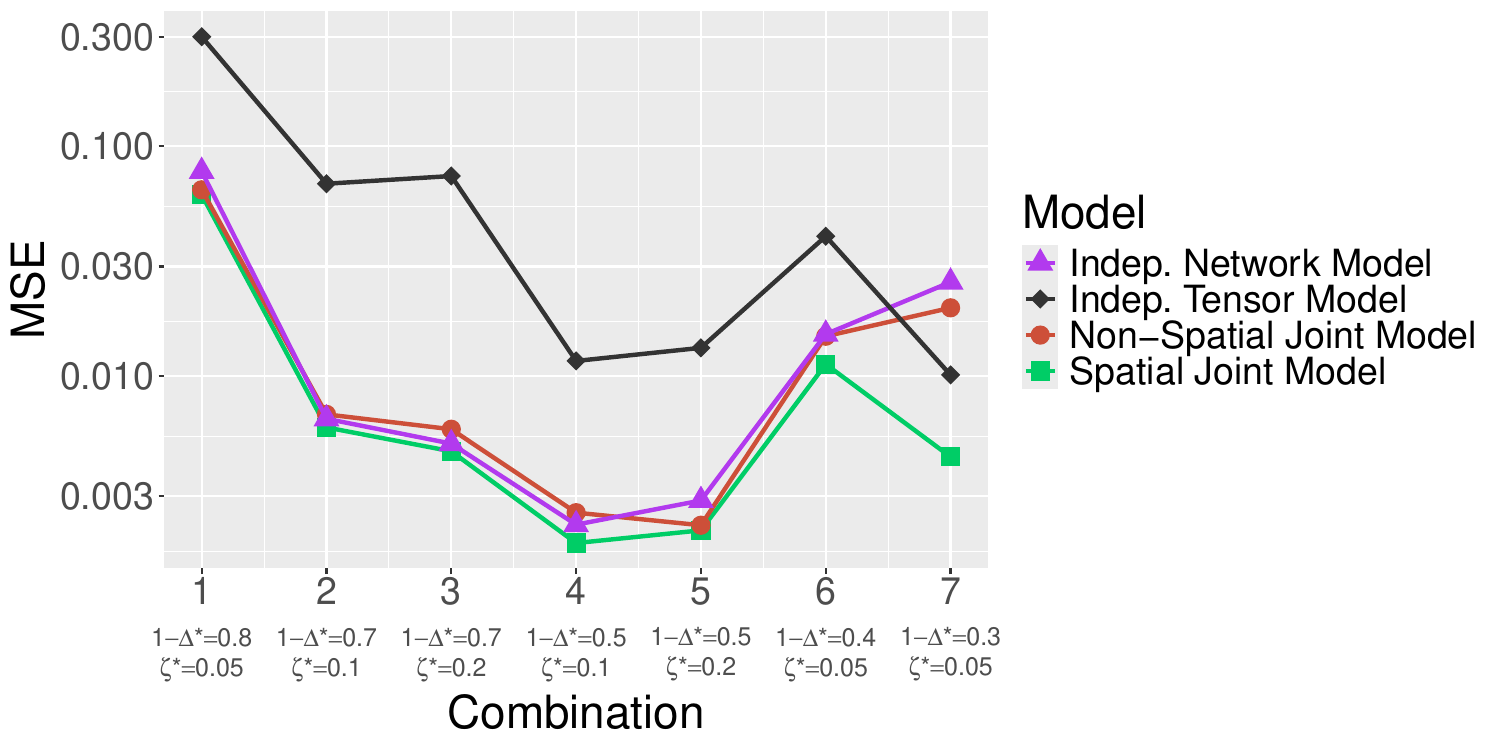}
    \label{mse-fig1}
}\hfill 
\subfloat[Scaled MSE of estimating the entries of $\balpha$]{%
    \includegraphics[width=0.49\linewidth]{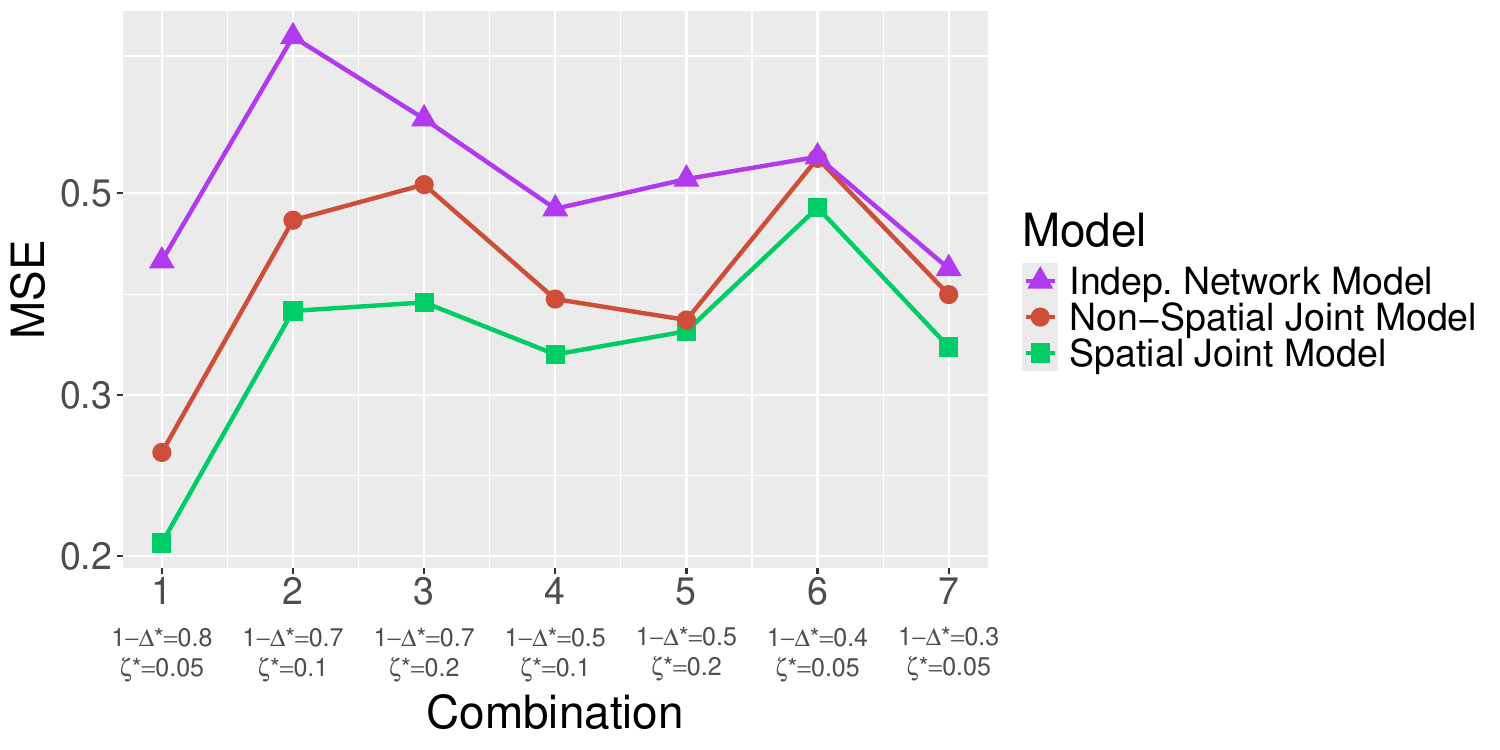}
    \label{mse-fig2}
}
\caption{Scaled MSE for estimating the predictor coefficients corresponding to the network outcome and nodal attributes across the seven simulation combinations. We present results for the proposed spatial joint model along with those of its competitors: independent tensor model, independent network model, and non-spatial joint model. Since the independent tensor model and independent network model fit the same model for nodal attributes, they will offer identical results in estimating $\balpha^*$. Hence, only results for the independent network model are presented in Figure \ref{mse-fig2}. The results show the overall superior performance of the spatial joint model in coefficient estimation.}
\label{mse-fig}
\end{figure}

Figure \ref{mse-fig} presents a comparative assessment of coefficient estimation for our method and its competitors. Figure \ref{mse-fig}(a) demonstrates that our method consistently achieves lower scaled MSE values when estimating 
$\bbeta^*$ across all simulation scenarios. Among competing methods, the independent tensor model performs the worst in most cases, except in Scenario 7, where node sparsity is at its lowest. This is likely due to its inability to account for the symmetric structure of the network. The independent network model performs comparably to our approach in high-sparsity settings (Scenarios 1, 2, and 3), but as sparsity decreases, it begins to lag behind the spatial joint model. The non-spatial joint model generally exhibits inferior performance, except in Scenarios 1 and 5, where node sparsity is at relatively high and mid-levels, respectively. These findings underscore the advantages of incorporating spatial correlation among nodes in the model. Notably, when node sparsity is low and spatial correlation is high (Scenarios 6 and 7), our approach significantly outperforms all competitors. Collectively, these comparisons highlight the benefits of our joint modeling framework, which effectively accounts for network structure and spatial correlation among nodes, leading to superior coefficient estimation, particularly under low-sparsity (e.g., $1 - \Delta^* = 0.3, 0.4$) and higher spatial correlations (e.g., $\zeta^*=0.05$).

Focusing on the point estimation of predictor coefficients for nodal attributes, Figure~\ref{mse-fig}(b) reports scaled MSE for each method across the seven simulation scenarios. The spatial joint model consistently achieves the lowest MSE among all competitors. Both spatial and non-spatial joint models show slight increase in MSE as node sparsity decreases. In contrast, the independent network model shows relatively stable but higher MSE across scenarios. These results highlight the advantages of incorporating spatial correlation through joint modeling, particularly under varying sparsity and spatial dependence.

\begin{figure}[h!]
    \centering
    \subfloat[Coverage of 95\% Credible Interval for $\bbeta$]{\includegraphics[width=0.49\linewidth]{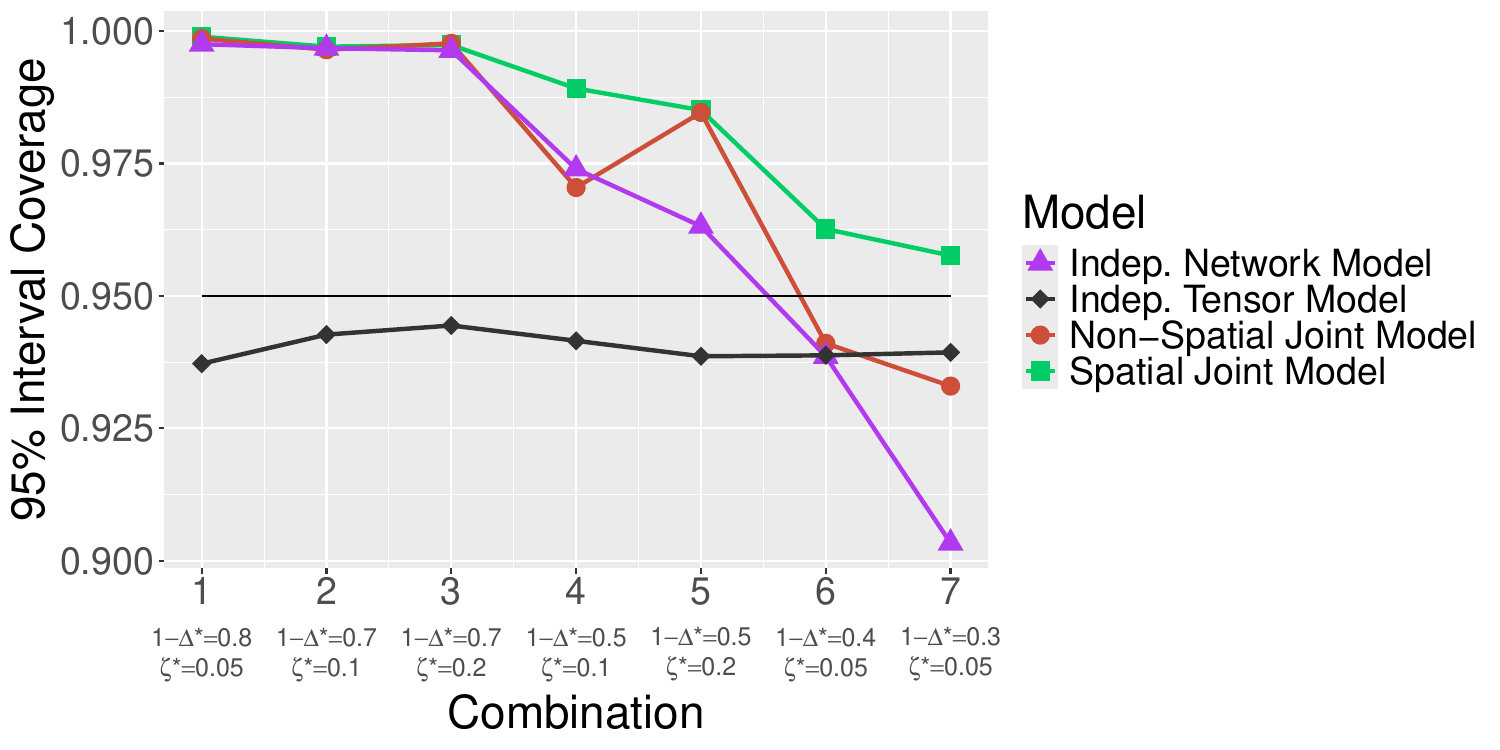}}
    \subfloat[Length of 95\% Credible Interval for $\bbeta$] {\includegraphics[width=0.49\linewidth]{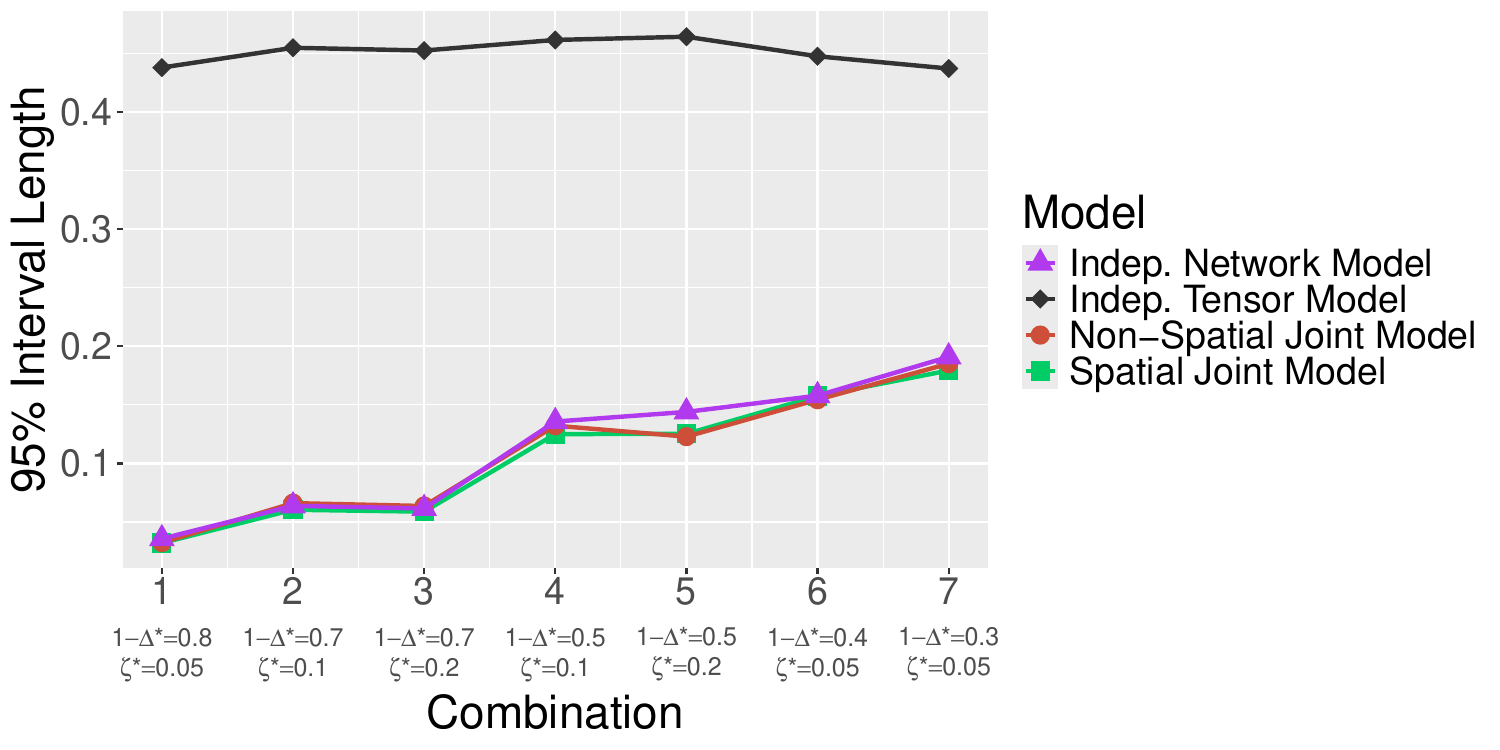}} \\
    \subfloat[Coverage of 95\% Credible Interval for $\balpha$]{\includegraphics[width=0.49\linewidth]{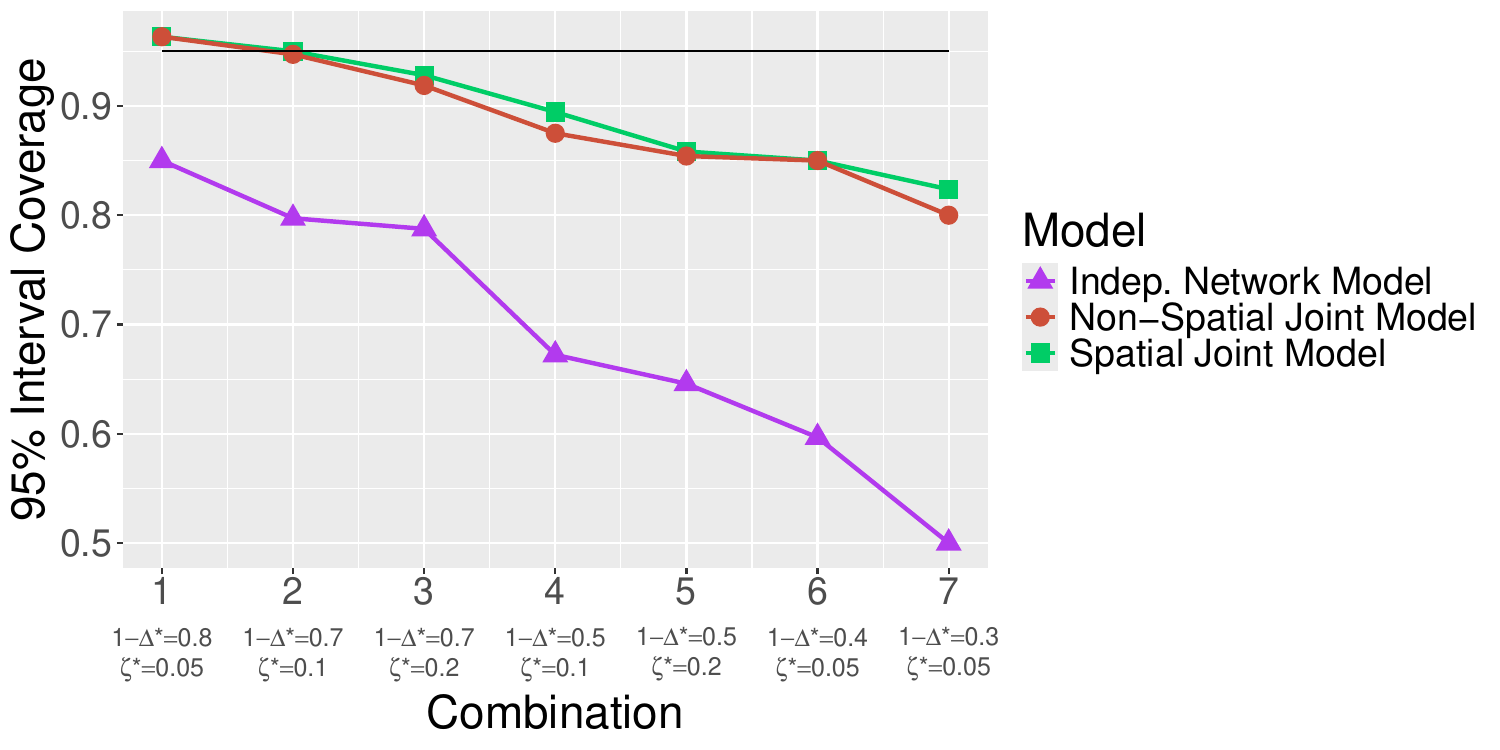}}
    \subfloat[Length of 95\% Credible Interval for $\balpha$]{\includegraphics[width=0.49\linewidth]{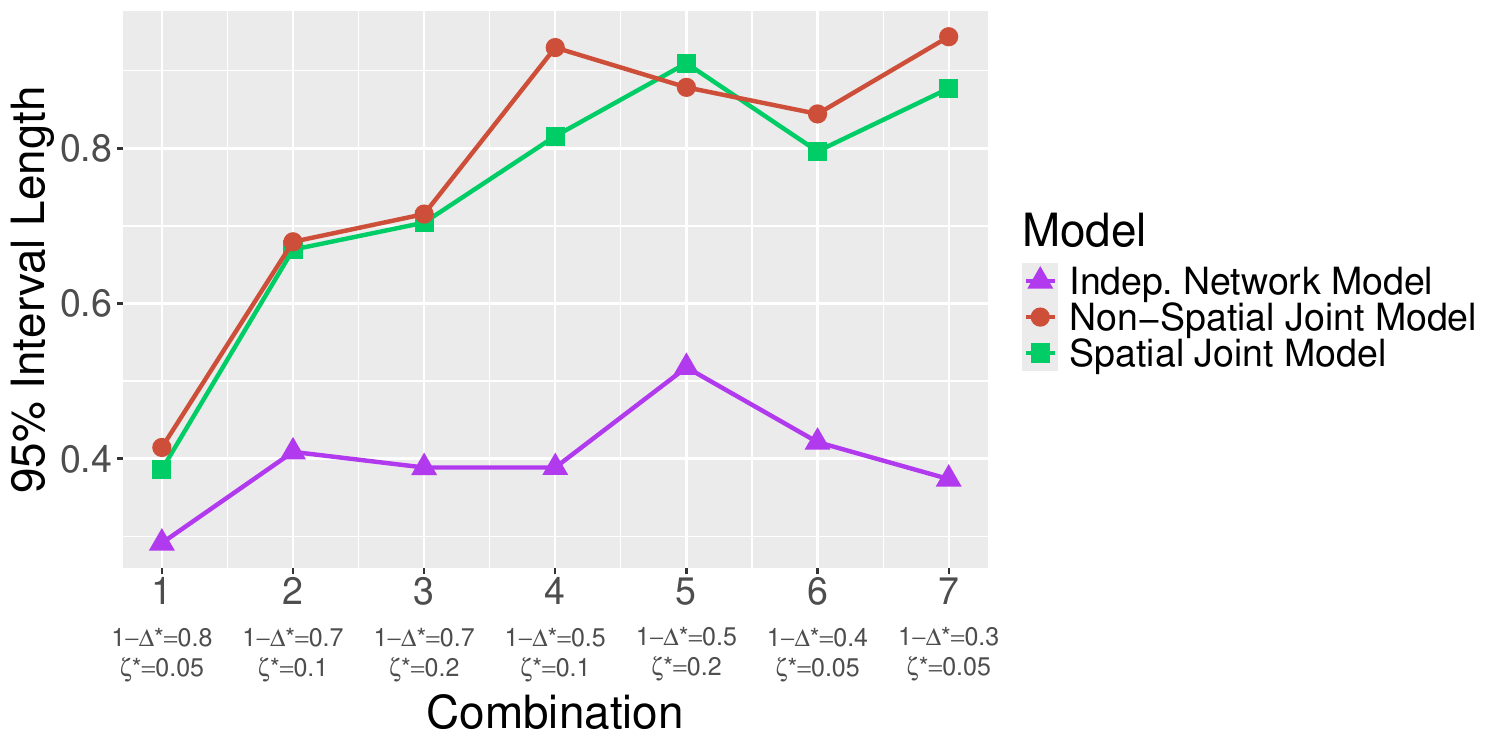}}
    \caption{Average length and coverage of 95\% credible intervals for Bayesian competitors and 95\% confidence intervals for frequentist competitors for $\bbeta$ and $\balpha$. The results show close to nominal coverage and overall shorter interval lengths for the spatial joint model, with most prominent advantages under scenarios 6 and 7, corresponding to low node sparsity and high spatial correlation.}\label{uncer}
\end{figure}

When evaluating the uncertainty quantification of the network coefficients in our spatial joint model, Figure \ref{uncer}(a) indicates that all interval coverages meet or exceed the nominal 95\% level. In contrast, the independent tensor model consistently exhibits interval coverage below 95\%. Additionally, both the independent network model and the non-spatial joint model demonstrate lower coverage in scenarios 6 and 7, where sparsity is at its lowest and spatial correlation is at its highest. The interval lengths, as shown in Figure \ref{uncer}(b), reveal that our model achieves these high coverage levels with intervals that are comparable in width, or even narrower, than those of competing approaches. This further underscores the advantages of joint modeling, which effectively accounts for network structure and incorporates spatial correlation among nodes.

For uncertainty quantification of $\balpha$, Figures \ref{uncer}(c) and \ref{uncer}(d) show the 95\% interval coverages and lengths across simulation scenarios. The spatial joint model achieves near-nominal coverage in high-sparsity settings (Scenarios 1-3), with coverage declining as sparsity decreases (Scenarios 4-7). The spatial joint model yields interval lengths comparable to the non-spatial joint model in scenarios 1-5, but produces narrower intervals when spatial correlation is high and node sparsity is low, due to its ability to incorporate spatial structure among nodal attributes. The independent network model produces narrower credible intervals for $\balpha$, consistent with its notable under-coverage. This stems from its failure to incorporate network information when estimating the relationship between nodal attributes and the predictor.

\subsubsection{Estimating Spatial Correlation Between Nodal Attributes}

\begin{figure}[h!]
\centering
\subfloat[Scenario 7: $1 - \Delta^* = 0.3$, $\zeta^* = 0.05$]{
\includegraphics[width=0.32\linewidth]{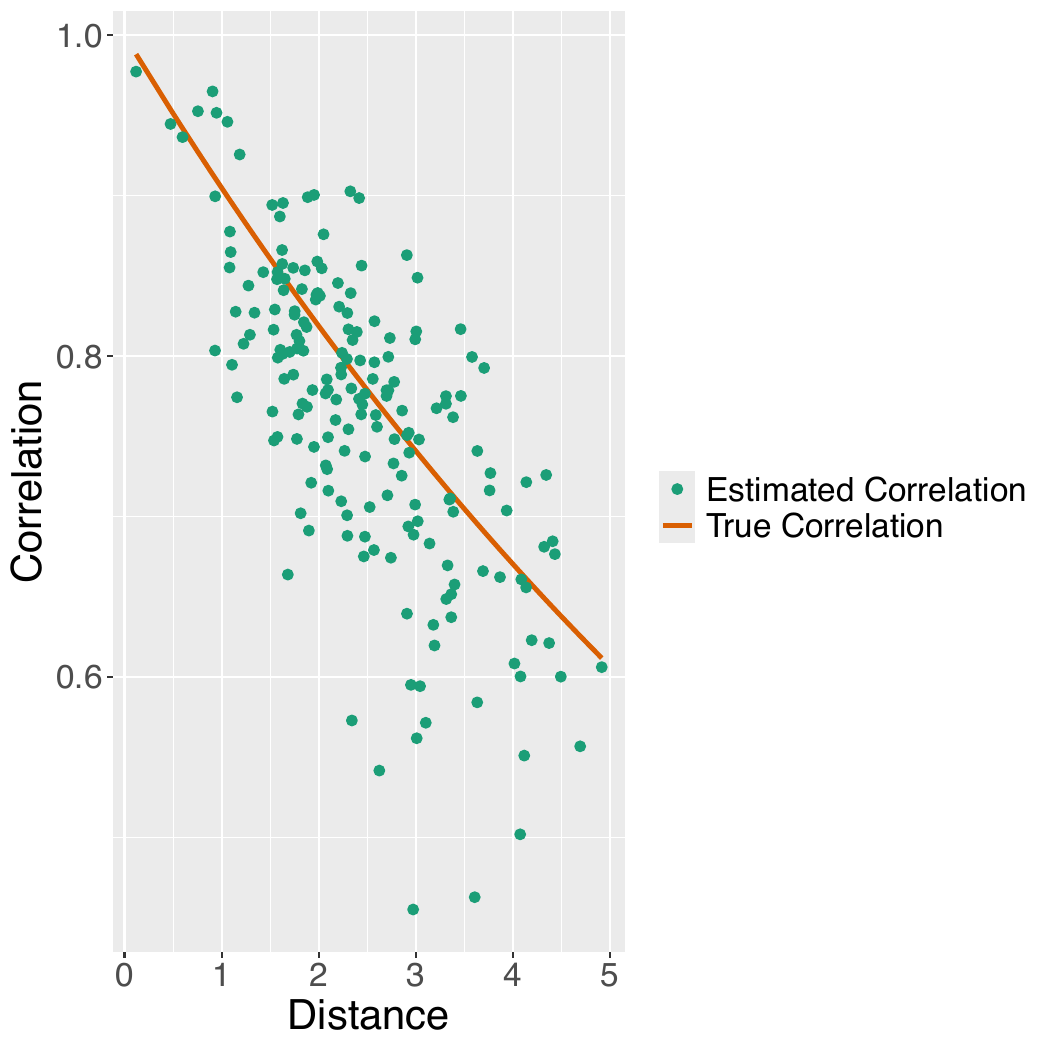}}
\subfloat[Scenario 2: $1 - \Delta^* = 0.7$, $\zeta^* = 0.1$]{
\includegraphics[width=0.32\linewidth]{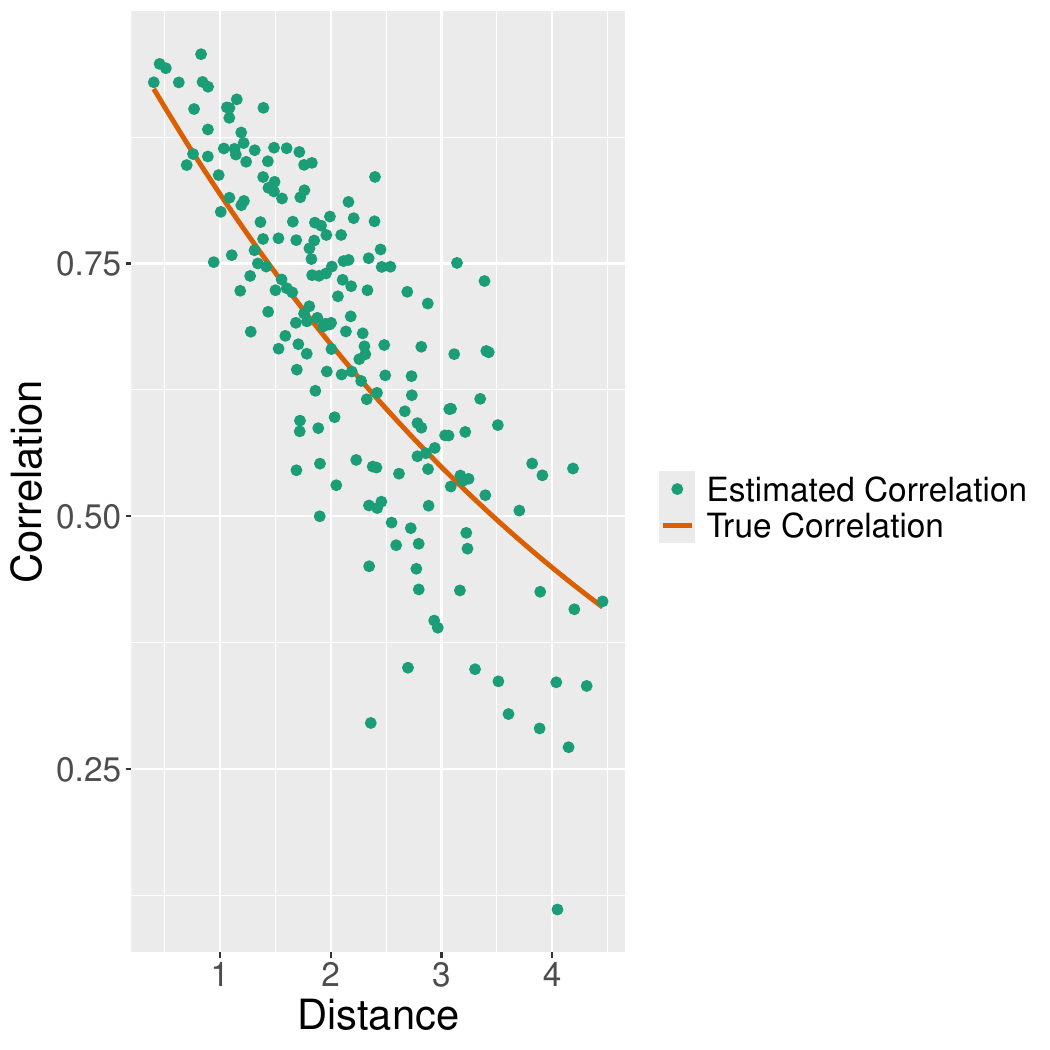}}
\subfloat[Scenario 5: $1 - \Delta^* = 0.5$, $\zeta^* = 0.2$]{
\includegraphics[width=0.32\linewidth]{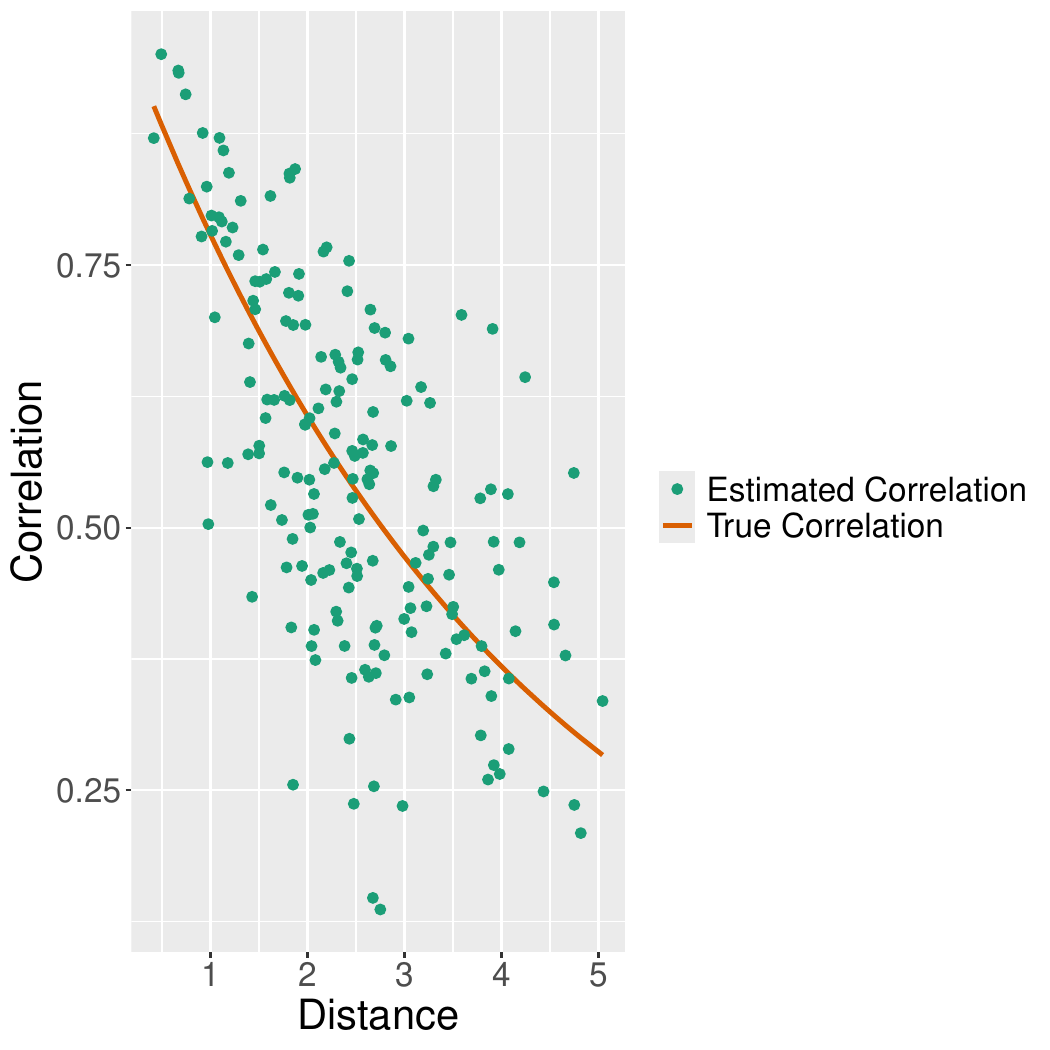}}
\caption{Each plot displays the spatial correlations estimated by our model against spatial distance, with the true exponential correlation function overlaid for reference. Results are shown for different values of the spatial scale parameter: $\zeta^* = 0.05, 0.1, 0.2$. Across all settings, the estimated correlations closely follow the underlying trend of the true spatial correlation as distance increases.}
\label{spatial-corr}
\end{figure}
Figure~\ref{spatial-corr} plots the estimated spatial correlation between any pair of locations and presents these correlations as a function of distance between the pair of locations. The true correlation function over distance is overlaid for reference. We present results for Scenarios 7, 2, and 5, corresponding to true spatial scale parameters $\zeta^*=0.05,0.1$, and $0.2$, respectively, representing a wide range of spatial scale. In all three cases, the estimated spatial correlations successfully capture the overall trend of the true correlations. For a smaller value of $\zeta^*=0.05$, where spatial correlation decreases more gradually as the distance between locations increases, the estimation process becomes more challenging. This results in a slight underestimation of spatial correlations at larger distances. However, for $\zeta^*=0.1,0.2$, this effect is negligible. Overall, a key strength of the proposed framework is its ability to incorporate spatial correlation between nodal attributes while simultaneously capturing their relationship with the network outcome, representing a novel contribution of our methodology.

\section{Analysis of Neuroimaging Data}\label{sec:realdata}

\begin{figure}[H]
\centering
\includegraphics[width=\textwidth]{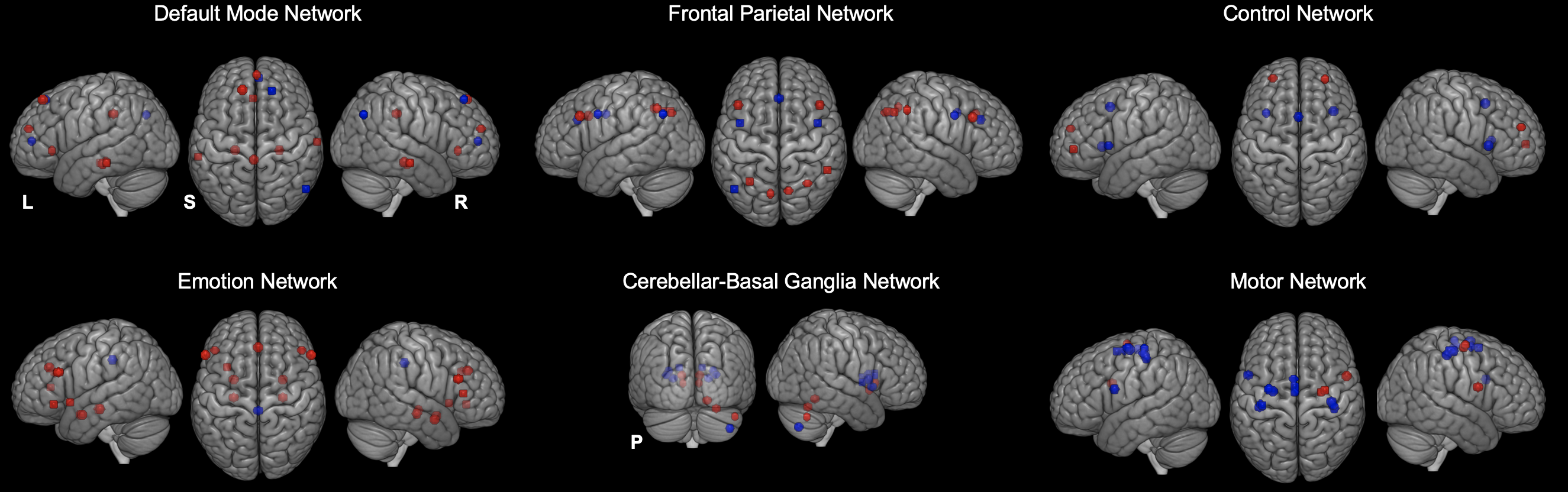}
\caption{The blue dots in the cortical and subcortical networks indicate the regions that are significantly associated with the Aggregate Pegboard Score.}
\label{real-data}
\end{figure}

The proposed spatial joint model is used to analyze the neuroimaging dataset described in Section \ref{sec:data_description}. Of the 69 regions of interest (ROIs), 33 regions are identified as significantly associated with the Aggregate Pegboard Score (APS) derived from the Purdue Pegboard Task. Figure~\ref{real-data} highlights cortical and subcortical ROIs that show significant associations with the behavioral measure. A list of the significant ROIs is provided in Table 1 of the supplementary material. 
The identified regions include several areas within the \emph{caudate and putamen}, \emph{motor cortical areas}, \emph{posterior cingulate cortex}, and \emph{frontal cortical regions}. This pattern is consistent with neurobiological expectations, as these regions are critically involved in motor planning, execution, and coordination - functions that are directly engaged by the Purdue Pegboard Task used in the analysis \citep{Picard2001}.

\emph{Motor cortical areas} are essential for executing motor movements like those required in the Purdue Pegboard Task, while the \emph{caudate and putamen} are established components of motor networks and play a key role in motor function \citep{DOYON200961, Robinson_2009, DiMartino_2008}. These subcortical structures are central to motor circuits and are known to be affected by aging and age-related neurodegenerative diseases \citep{SEIDLER2010721,  Bo_2014, Bell_2014}. Their presence in our findings supports current understanding of  motor system organization and age-related changes, further demonstrating the relevance and utility of our novel modeling approach.
More broadly, as illustrated in Figure~\ref{real-data}, both the \emph{motor and cerebellar-basal ganglia networks} emerge as key components identified by our analysis - consistent with their well-established roles in motor behavior and coordination.
In parallel, the involvement of \emph{frontal regions} aligns with expectations for tasks engaging cognitive functions \citep{Courtney_1998}. These nodes span both the frontal parietal and cognitive control networks. Although the Aggregate Pegboard Score combines motor and cognitive components, the assembly task in the Purdue Pegboard Test likely draws on working memory to sequence actions, which corresponds to the activation of \emph{prefrontal regions} highlighted by our model. These findings are consistent with the more prominent involvement of associated cognitive networks.

Interestingly, several nodes within the \emph{default mode network} are also implicated, primarily within \emph{frontal and parietal} cortical areas. While the \emph{default mode network} is traditionally associated with self-directed thought and is typically ``turned off" during task performance, it has nonetheless been linked to cognitive functioning in aging populations \citep{ANDREWSHANNA2007924}. Its involvement here underscores its broader role in supporting behavioral performance, even in task-focused settings. Finally, only one region in the \emph{emotion network} is found to significantly predict behavior when considering both structural and functional data. This limited contribution highlights the specificity of our proposed approach. Since the Purdue Pegboard Task primarily engages motor and cognitive systems rather than emotional processing, we would not expect broad involvement of emotion-related regions. The fact that these areas are largely excluded from the model’s selected predictors suggests that our method accurately identifies brain regions most relevant to the task, while effectively filtering out those not functionally aligned with motor and cognitive performance. Thus, by employing the proposed approach, the model identifies a set of brain regions that align closely with established literature on motor and cognitive function as well as current knowledge of functional neuroanatomy. Importantly, it achieves this with functional specificity while providing principled uncertainty quantification.

\begin{figure}[H]
\centering
\includegraphics[width=\linewidth]{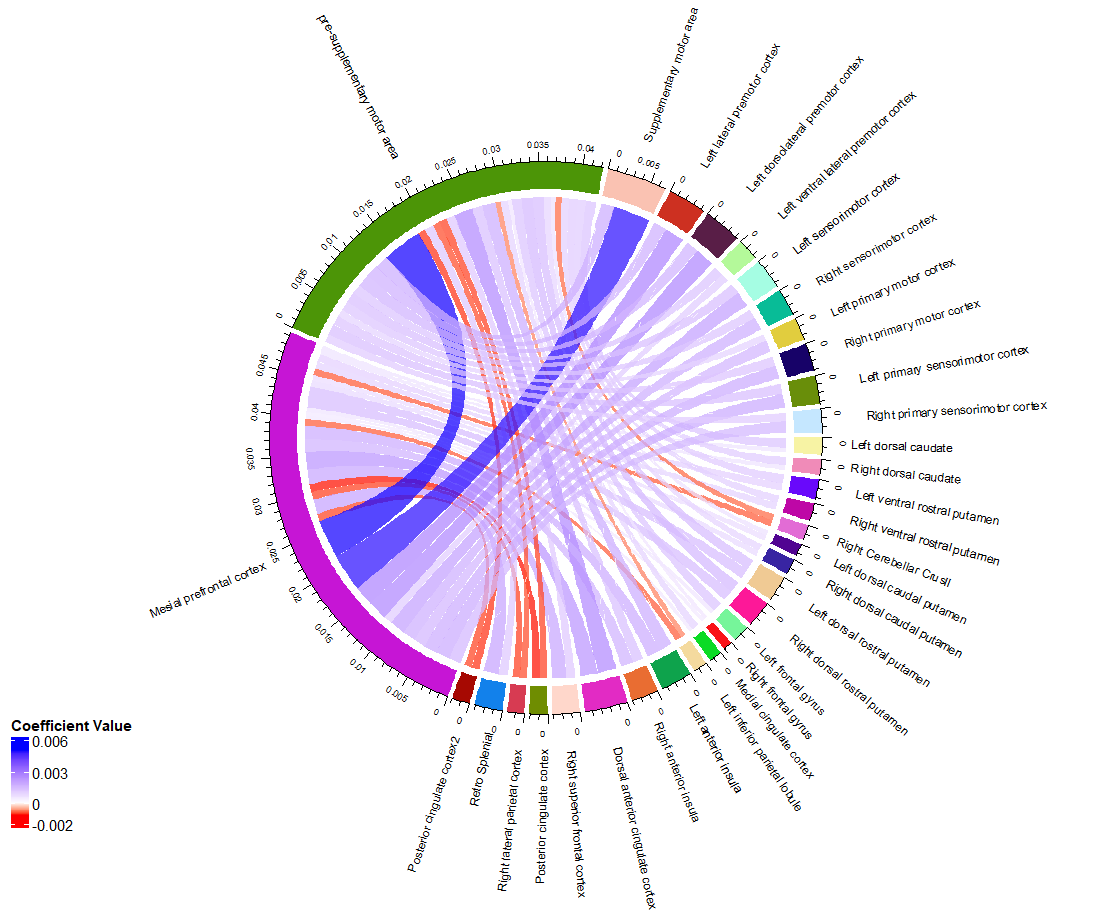}
\caption{Circos plot depicting the estimated network coefficients among the 33 ROIs identified as influential.}
\label{real-data_coefficient}
\end{figure}
Focusing now on the estimation of network coefficients, we note that our model and prior formulation in equations (\ref{param}) and (\ref{prior_u}) implies $\beta(u,v) = 0$ whenever either the $u$th or $v$th ROI is deemed uninfluential. The estimated nonzero network coefficients $\beta(u,v)$ between influential ROIs are visualized in the circos plot in Figure~\ref{real-data_coefficient}. 
The plot highlights strong associations between the Aggregate Pegboard Score (APS) and the interconnections between the mesial prefrontal cortex and a subset of ROIs. The connections between the pre-supplementary motor area also exhibit prominent relationships with APS. 

\subsection{Predictive Inference}
Although predictive inference for network and nodal attributes is not the primary focus of this study, we report results for predicting both the edges between ROIs in the connectivity network and the structural attributes at the ROIs to assess the performance of the competing models. Table~\ref{tab:real_data_pred} shows that the spatial joint model, non-spatial joint model, and independent network model all deliver competitive performance, while the independent tensor model yields less accurate point prediction and significant under-coverage due to quite narrow predictive intervals.  Moreover, the spatial joint model achieves nominal coverage, while the others exhibit mild under-coverage. Overall, the spatial joint model emerges as a robust tool for inference on influential nodes, regression coefficients, spatial dependencies among nodal attributes, and predictive outcomes.

\begin{table}[H]
    \centering
        \begin{tabular}{|c|c|c|c|c|c|c|}
        \hline
        & \multicolumn{3}{|c|}{fMRI Network Edges} & \multicolumn{3}{|c|}{Nodal Attributes} \\
        \hline
        Model & MSPE & Coverage & Length & MSPE & Coverage & Length\\
        \hline
         Spatial Joint Model & 0.9971 & 0.9417 & 4.0025 & 0.8152 & 0.9515 & 3.6091 \\
         Non-Spatial Joint Model & 1.0019 & 0.9347 & 3.8836 & 0.8202 & 0.9387 & 3.4832 \\
         Independent Network Model & 1.0037 & 0.9316 & 3.8382 & 0.8204 & 0.9310 & 3.3615\\
         Independent Tensor Model & 1.1006 & 0.3200 & 0.8391  & --- & --- & --- \\
         \hline
    \end{tabular}
    \caption{MSPE and 95\% prediction interval coverages and lengths for prediction of the fMRI network edges and structural data. The prediction results of the Independent Tensor Model on the nodal attributes are the same as those of the Independent Network Model.}
    \label{tab:real_data_pred}
\end{table}

\section{Conclusion and Future Work}\label{sec:conclusion}
This article is motivated by resting-state functional MRI (fMRI) and structural MRI (sMRI) data collected in the Cognitive and Motor Aging Neuroimaging Laboratory at Texas A\&M University, with the scientific aim of identifying brain regions significantly associated with cognitive and motor aging. A brain connectome is constructed from the fMRI data, where ROIs serve as network nodes, and sMRI measurements at each ROI are treated as node-specific attributes. A composite measure of cognitive and motor function, the Aggregate Pegboard Score (APS), is used as a subject-level predictor in the analysis. To jointly analyze these data, we develop a predictor-dependent modeling framework that jointly learns the structure of the brain network and nodal attributes. The proposed model enables inference on influential brain regions related to cognitive and motor aging and estimates the effect of predictors on both network and nodal attributes. Spatial information on ROIs is explicitly incorporated to capture spatially varying associations in nodal attributes. The empirical results indicate that the new method can offer improved performance over existing alternatives, highlighting the benefits of integrating network topology, joint modeling of network and nodal attributes, and spatial correlation among nodal attributes over ROIs. Our framework identifies scientifically meaningful ROIs associated with cognitive and motor aging.
Future work will be directed toward modeling more complex, non-linear dependencies among brain networks, nodal attributes, and subject-level covariates. We propose the development of an interpretable deep learning framework capable of estimating the underlying regression function while retaining the interpretability crucial for identifying influential network nodes.

\section*{Acknowledgments}
Portions of this research were conducted with the advanced computing resources provided by Texas A\&M Department of Statistics Arseven Computing Cluster.

\bibliography{References}

\end{document}


\maketitle

\abstract{This supplementary material consists of two sections. Section 1 outlines the full conditional distributions of the model parameters used for Gibbs sampling. Section 2 contains the list of regions of interest (ROIs) identified by the proposed spatial joint model as being significantly associated with the Aggregate Pegboard Score.} 

\newpage

\section{Full Conditional Distributions} The following full conditional distributions are presented for the case of $p = 1$.  Let $\by_{i} = (y_{i}(u, v) : 1 \leq u < v \leq V)^{T}$ be the vector of upper triangular entries of the $i$-th network adjacency matrix, $\bY_{i}$, and $\widetilde{\bbeta} = (\beta(u, v) : 1 \leq u < v \leq V)^{T}$ be the vector of upper triangular entries of the network adjacency matrix $\bB$, both of size $h = V(V-1)/2$. Furthermore, let $\widetilde{\balpha} = (\alpha(1),...,\alpha(V))^T$, $\bz_i = (z_i(\bs_1),...,z_i(\bs_V))^T$, $\bgamma_{y} = (\gamma_{1,y},...,\gamma_{q,y})^T$, and $\bgamma_{z} = (\gamma_{1,z},...,\gamma_{q,z})^T$. Lastly, let $\bSigma \in \mathbb{R}^{V \times V}$ be a matrix whose $(u, 
 v)$-th entry is $\exp(-\zeta||\bs_u-\bs_v||)$. Then,

\begin{itemize}
    \item $\mu_y | - \sim N(\dfrac{1}{nh}\textbf{1}_{h}^{T} \sum_{i=1}^{n} (\by_{i} - \widetilde{\bbeta} x_{i} - \textbf{1}_{h} \bw_{i}^{T} \bgamma_{y}), \dfrac{\tau_{y}^{2}}{nh})$.

    \item $\mu_{z} | - \sim N \left(\dfrac{1}{n \textbf{1}_{V}^{T} \bSigma^{-1}\textbf{1}_{V}} \textbf{1}_{V}^{T} \bSigma^{-1} \sum_{i = 1}^{n} (\bz_{i}  -\widetilde{\balpha} x_{i} - \textbf{1}_V\bw_{i}^{T} \bgamma_{z}), \dfrac{\tau_{z}^{2}}{n \textbf{1}_{V}^{T} \bSigma^{-1}\textbf{1}_{V}}\right)$.

    \item $ \bgamma_{y} \mid - \sim N(\bmu_{\bgamma_{y}},\bSigma_{\bgamma_{y})} $, 
    
\noindent where $ \bSigma_{\bgamma_{y}} = \dfrac{\tau_{y}^{2}}{h} (\sum_{i=1}^{n} \bw_{i}  \bw_{i}^{T})^{-1}  $ and $\bmu_{\bgamma_{y}} = \bSigma_{\bgamma_{y}} \frac{1}{\tau_y^2}  \sum_{i=1}^{n} \bw_{i} \textbf{1}_{h}^{T}(\by_{i} - \textbf{1}_{h} \mu_{y} - \widetilde{\bbeta} x_{i})$.

\item $ \bgamma_{z} \mid - \sim N(\bmu_{\bgamma_{z}},\bSigma_{\bgamma_{z})} $, 

\noindent where $ \bSigma_{\bgamma_{z}} = \dfrac{\tau_{z}^{2}}{\textbf{1}_V^T\bSigma^{-1}\textbf{1}_V} (\sum_{i=1}^{n} \bw_{i}  \bw_{i}^{T})^{-1} $ and $\bmu_{\bgamma_{z}} = \bSigma_{\bgamma_{z}} \frac{1}{\tau_z^2} \sum_{i=1}^{n} \bw_{i}\textbf{1}_V^T\bSigma^{-1} (\bz_{i} - \textbf{1}_{V} \mu_{z} - \widetilde{\balpha} x_{i})$.

\item $ \tau_{y}^{2} \mid - \sim
IG(a+\dfrac{hn}{2}, b + \dfrac{1}{2} \sum_{i=1}^{n}
(\by_{i} - \textbf{1}_{h} \mu_{y} - \widetilde{\bbeta} x_{i} - \textbf{1}_{h} \bw_{i}^{T} \bgamma_{y})^{T} (\by_{i} - \textbf{1}_{h} \mu_{y} - \widetilde{\bbeta} x_{i} - \textbf{1}_{h} \bw_{i}^{T} \bgamma_{y}))$.

\item $ \tau_{z}^{2} | - \sim IG\left(a + \dfrac{n V}{2}, b + \dfrac{1}{2} \sum_{i = 1}^{n} 
(\bz_{i} - \textbf{1}_{V} \mu_{z} - \widetilde{\balpha} x_{i} -  \textbf{1}_V\bw_{i}^{T} \bgamma_{z})^T\bSigma^{-1}(\bz_{i} - \textbf{1}_{V} \mu_{z} - \widetilde{\balpha} x_{i} -  \textbf{1}_V\bw_{i}^{T} \bgamma_{z})\right)$.

\item $\Delta \mid - \sim Beta(a_{\Delta} + \sum_{v=1}^{V} \eta_{v}, b_{\Delta} + V - \sum_{v=1}^{V} \eta_{v})$.

\item $\bL | - \sim IW(\nu+\#\{ v : \eta_{v} = 1 \}, \bSigma_{\bL} + \sum_{v : \eta_{v} = 1}\bxi_v\bxi_v^{T})$,

\noindent where $\bxi_v = (\alpha(v),\btheta(v)^T)^T$.

\item $(\pi_{1,r}, \pi_{2,r}, \pi_{3,r}) \mid - \sim 
Dirichlet(r^{\xi} + I(\lambda_{r} = 0), 1 +  I(\lambda_{r} = 1), 1 + I(\lambda_{r} = -1))$.

\item $\lambda_{r} \mid - \sim 
\begin{cases}
      0 & w.p. \; p_{1,r}, \\
      1 & w.p. \; p_{2,r}, \\
      -1 & w.p. \; p_{3,r}, 
\end{cases}$

\noindent where $p_{1,r} = \dfrac{\pi_{1, r} \prod_{i = 1}^{n} N(\by_{i} \mid \textbf{1}_{h}  \mu_y + \widetilde{\bbeta}^{(\lambda_{r} = 0)} x_{i} + \textbf{1}_{h} \bw_{i}^{T} \bgamma_{y}, \tau_{y}^{2} \bI_{h})}{S} $, 

\noindent $p_{2,r} = \dfrac{\pi_{2, r} \prod_{i = 1}^{n} N(\by_{i} \mid \textbf{1}_{h}  \mu_y + \widetilde{\bbeta}^{(\lambda_{r} = 1)} x_{i} + \textbf{1}_{h} \bw_{i}^{T} \bgamma_{y}, \tau_{y}^{2} \bI_{h})}{S} $,

\noindent $p_{3,r} = \dfrac{\pi_{3, r} \prod_{i = 1}^{n} N(\by_{i} \mid \textbf{1}_{h}  \mu_y + \widetilde{\bbeta}^{(\lambda_{r} = -1)} x_{i} + \textbf{1}_{h} \bw_{i}^{T} \bgamma_{y}, \tau_{y}^{2} \bI_{h})}{S} $,

\noindent and 
$S = \pi_{1, r} \prod_{i = 1}^{n} N(\by_{i} \mid \textbf{1}_{h}  \mu_y + \widetilde{\bbeta}^{(\lambda_{r} = 0)} x_{i} + \textbf{1}_{h} \bw_{i}^{T} \bgamma_{y}, \tau_{y}^{2} \bI_{h})$

$\;\;\;\;+ \pi_{2, r} \prod_{i = 1}^{n} N(\by_{i} \mid \textbf{1}_{h}  \mu_y + \widetilde{\bbeta}^{(\lambda_{r} = 1)} x_{i} + \textbf{1}_{h} \bw_{i}^{T} \bgamma_{y}, \tau_{y}^{2} \bI_{h})$

$\;\;\;\; + \pi_{3, r} \prod_{i = 1}^{n} N(\by_{i} \mid \textbf{1}_{h}  \mu_y + \widetilde{\bbeta}^{(\lambda_{r} = -1)} x_{i} + \textbf{1}_{h} \bw_{i}^{T} \bgamma_{y}, \tau_{y}^{2} \bI_{h})$.

\noindent Here, $ \widetilde{\bbeta}^{(\lambda_{r} = c)}$  is the vector, $\widetilde{\bbeta}$, obtained when setting $\lambda_{r} = c$, in $\bLambda$, where $c \in \{-1, 0, 1\}$.
\\

\item For the update of $ \bxi_v = 
     (\alpha(v),
      \btheta(v)^T)^T$,       
let $
\by_{i}^{(v)} = (y_i(1,v),...,y_i(v-1,v), y_i(v,v+1),... y_i(v,V))^T \in \mathbb{R}^{V-1}
$, with 
$
\widetilde{\by}_{i}^{(v)} = \by_{i}^{(v)} - \mu_{y} \textbf{1}_{V - 1} - \bgamma_{y}^{T} \bw_i \textbf{1}_{V - 1}$
and $\widetilde{z}_{i}(\bs_v) = z_{i}(\bs_v) - \mu_{z} - \bgamma_{z}^{T} \bw_i$.

\noindent Furthermore, define
$
\btheta^{(-v)} = (\btheta(1), ..., \btheta(v-1), \btheta(v+1),...,\btheta(V))^T \in \mathbb{R}^{(V-1) \times R}
$
and $\be_{i}^{(v)} \overset{iid}{\sim} N(\bzero_{V-1}, \tau_{y}^{2} \bI_{V-1})
$.

\noindent Moreover, let $
\boldf_i^{(v)} = \begin{bmatrix}
     \widetilde{z}_{i}(\bs_v) \\
     \widetilde{\by}_{i}^{(v)}
\end{bmatrix} \in \mathbb{R}^{V}
$ and $\bA_{i} = \begin{bmatrix}
x_i & \bzero^T_{R} \\
\bzero_{V-1} & x_i \btheta^{(-v)} \bLambda
\end{bmatrix}  \in \mathbb{R}^{V \times (R+1)}
$.

\noindent Lastly, let
$
\boldf^{(v)} = \begin{bmatrix}
\boldf_{1}^{(v)} \\
\vdots \\
\boldf_{n}^{(v)}
\end{bmatrix} \in \mathbb{R}^{nV}$ and 
$
\bA = \begin{bmatrix}
\bA_{1} \\
\vdots \\
\bA_{n}
\end{bmatrix} \in \mathbb{R}^{nV \times (R+1)}
$. 
 
\noindent Then $
\boldf^{(v)} = \bA \bxi_{v} + \widetilde{\be}^{(v)}$, where 
$\widetilde{\be}^{(v)} \sim N(\bzero_{nV}, \bSigma_{\boldf})$, with $\bSigma_{\boldf} = \bI_{n} \otimes \begin{bmatrix}
     \tau_{z}^{2} & \bzero^T_{V-1} \\
     \bzero_{V-1} &
     \tau_{y}^{2} \bI_{V-1} 
\end{bmatrix}$. 

\noindent Hence,

$$\bxi_v \mid - \sim \eta_v N(\bmu_{\bxi_{v}}, \bSigma_{\bxi_{v}}) + (1-\eta_{v}) \delta_{0},$$

\noindent where 
$
\bSigma_{\bxi_{v}} = (\bA^T \bSigma_{\boldf}^{-1} \bA + \bL^{-1})^{-1}
$ and
$
\bmu_{\bxi_{v}} = \bSigma_{\bxi_{v}} \bA^{T} \bSigma_{\boldf}^{-1} \boldf^{(v)}$.

\item $
\eta_{v} \mid - \sim Ber(\Delta_{\eta_{v}})
$, 
\noindent where $\Delta_{\eta_{v}} = \dfrac{\Delta N(\boldf^{(v)} \mid  \bzero_{nV}, \bA \bL \bA^{T} + \bSigma_{\boldf})}{\Delta N(\boldf^{(v)} \mid  \bzero_{nV}, \bA \bL \bA^{T} + \bSigma_{\boldf}) + (1 - \Delta) N(\boldf^{(v)} \mid \bzero_{nV}, \bSigma_{\boldf})} $.

\item For $\zeta$, over the grid, $\zeta_1,...,\zeta_g$, 

\noindent $P(\zeta=\zeta_l|-)=\frac{\prod_{i=1}^n N(\bz_i| \textbf{1}_V \mu_{z} + \widetilde{\balpha} x_i+\textbf{1}_V \bw_i^T \bgamma_z,\tau_z^2\bSigma(\zeta_l))}{\sum_{m=1}^g\prod_{i=1}^n N(\bz_i| \textbf{1}_V \mu_{z} + \widetilde{\balpha} x_i+\textbf{1}_V\bw_i^T\bgamma_z,\tau_z^2\bSigma(\zeta_m))},\:\:l=1,...,g,$

\noindent where $\bSigma(\zeta_l)$ is the resulting $\bSigma$ matrix when $ \zeta = \zeta_l$.

\end{itemize}

\newpage

\section{Selected Regions of Interest (ROIs)}  Table \ref{selected_roi_table} lists the 33 ROIs identified by the proposed spatial joint model as being significantly associated with the Aggregate Pegboard Score, along with the cortical or subcortical networks they belong to.

\begin{table}[h]
    \centering
    \setlength{\extrarowheight}{-9pt}
    \begin{tabular}{|c|c|}
    \hline
         ROI & Cortical/Subcortical Network \\
    \hline
         Left dorsal caudate & Cerebellar-basal ganglia network \\
         Right dorsal caudate & Cerebellar-basal ganglia network  \\
         Left ventral rostral putamen & Cerebellar-basal ganglia network \\
         Right ventral rostral putamen & Cerebellar-basal ganglia network  \\
         Right cerebellar crusII & Cerebellar-basal ganglia network \\
         Left dorsal caudal putamen & Cerebellar-basal ganglia network \\
         Right dorsal caudal putamen & Cerebellar-basal ganglia network \\
         Left dorsal rostral putamen & Cerebellar-basal ganglia network \\
         Right dorsal rostral putamen & Cerebellar-basal ganglia network \\
         Left frontal gyrus & Frontal-parietal network \\
         Right frontal gyrus & Frontal-parietal network \\
         Medial cingulate cortex & Frontal-parietal network \\
         Left inferior parietal lobule & Frontal-parietal network \\
         Left anterior insula & Control network \\
         Right anterior insula & Control network \\
         Dorsal anterior cingulate cortex & Control network \\
         Right superior frontal cortex & Default mode network \\
         Posterior cingulate cortex & Default mode network \\
         Right lateral parietal cortex & Default mode network \\
         Retro Splenial & Default mode network \\
         Posterior cingulate cortex2 & Emotion network \\
         Mesial prefrontal cortex & Motor network \\
         Pre-supplementary motor area & Motor network \\
         Supplementary motor area & Motor network \\
         Left lateral premotor cortex & Motor network \\
         Left dorsolateral premotor cortex & Motor network \\
         Left ventral lateral premotor cortex & Motor network \\
         Left sensorimotor cortex & Motor network \\
         Right sensorimotor cortex & Motor network \\
         Left primary motor cortex & Motor network \\
         Right primary motor cortex & Motor network \\
         Left primary sensorimotor cortex & Motor network
 \\
         Right primary sensorimotor cortex & Motor network \\
    \hline
    \end{tabular}
    \caption{List of the 33 ROIs selected by the proposed spatial joint model and their associated cortical or subcortical networks.}
    \label{selected_roi_table}
\end{table}